\documentclass[article,groupedaddress]{revtex4-2}
\usepackage[T2A]{fontenc}
\usepackage[utf8]{inputenc}
\usepackage[russian, english]{babel}

\usepackage[nottoc]{tocbibind}
\usepackage{amsmath}
\usepackage{braket}
\usepackage{amssymb}
\usepackage{graphicx}
\usepackage{placeins}

\usepackage{booktabs}
\usepackage{multirow}
\usepackage{setspace}

\usepackage{enumitem}
\usepackage{listings}
\usepackage{colortbl}
\usepackage[outline]{contour}
\usepackage{bm}
\usepackage{xparse}

\usepackage{amsfonts}
\usepackage{titlesec}
\usepackage{wasysym}

\usepackage{epstopdf}
\usepackage{xcolor, soul} 
\definecolor{Gray}{gray}{0.85}
\newcolumntype{g}{>{\columncolor{Gray}}l}
\usepackage[normalem]{ulem} 

\usepackage{dsfont}
\usepackage{bbold}

\begin{document}

	\title{Three-Path Quantum Cheshire Cat Observed in Neutron Interferometry}

	\author{Armin Danner$^{1}$}
	\email{armin.danner@tuwien.ac.at}
	\author{Niels Geerits$^1$}
	\author{Hartmut Lemmel$^{1,2}$}
	\author{Richard Wagner$^1$}
	\author{Stephan Sponar$^1$}
	\author{Yuji Hasegawa$^{1,3}$}
	\email{yuji.hasegawa@tuwien.ac.at}
	\affiliation{%
		$^1$Atominstitut, TU Wien, Stadionallee 2, 1020 Vienna, Austria \\
		$^2$Institut Laue-Langevin, 38000 Grenoble, France\\
		$^3$Department of Applied Physics, Hokkaido University, Kita-ku, Sapporo 060-8628, Japan
	}

	\date{\today}
	
	\hyphenpenalty=800\relax
	\exhyphenpenalty=800\relax
	\sloppy
	\setlength{\parindent}{0pt}
	
	\noindent
	
	\begin{abstract}
		The paradoxical phenomenon of the quantum Cheshire Cat (qCC) refers to situations where different properties of a particle appear to be localised in different paths of an interferometer and therefore spatially separated. This observation is obtained by implementing a pre- and postselection procedure. The localisations are determined qualitatively through conspicuous changes induced by weak interactions. Previous demonstrations of the qCC only used the path and spin/polarisation degrees of freedom. In addition, the present experiment uses the neutron's energy as a third property in a three-path interferometer. 
		It is demonstrated that the three properties of neutrons are found separated in different paths in the interferometer; a detailed analysis suggests that the appearance of a property is strongly related to the geometrical relation between the state vectors of pre- and postselection with weak interactions in between. If a weak interaction in a path locally generates a state vector with a component parallel to the reference state in another path, a conspicuous intensity oscillation is expected and observed. 
		Therefore, the appearance of the observed intensity oscillations is attributed solely to the cross-terms between the reference and the newly generated state via weak interactions. 
	\end{abstract}
	
	\maketitle
	
	\section{Introduction}
	Since the introduction of quantum mechanics, its theoretical framework has suggested counter-intuitive and paradoxical phenomena: the quantisation of energy transfer in the treatment of black body radiation \cite{Planck1901}, wave-particle duality \cite{deBroglie1924} and delayed choice \cite{WHEELER19789, Wagner2023} are only three of the most popular ones. Their study provides us with a deeper understanding of nature and opportunities for new technology.\cite{Ekert1991, Ourjoumtsev2006, Nielsen2010, Neamen2012} 
	\\
	Usually, a particle and its properties are considered as inseparable. In contrast, Aharonov \textit{et al}.\ \cite{Aharonov_2013} described intriguing interferometer experiments in which different properties of a physical entity appear to be spatially separated -- localised in different sub-beams of an interferometer. Aharonov \textit{et al.} coined the term quantum Cheshire Cat (qCC) in tribute to similar behaviour of the so-called Cheshire Cat in Lewis Carroll's ‘‘Alice's Adventures in Wonderland’’ \cite{Carroll1866}. In the novel, different parts of the Cheshire Cat can appear independently of each other. The entity which will be investigated in the present article is the neutron with its properties of spin, particle and energy which will appear to be separated into three different paths of an interferometer. 
	\\
	The apparent separation of properties emerges from the applied pre- and postselection procedure: all sub-states of the sub-beams in the interferometer are preselected/prepared to be mutually orthogonal when being recombined, and the final postselection allows only one sub-beam -- the reference beam -- to be transmitted towards the detector. If an additional interaction in one of the other paths midway between pre- and postselection creates a component parallel to the reference, an intensity oscillation will appear at the output of the interferometer which depends on the relative phase. This is the conventional explanation of such intensity oscillations as an interference effect through the cross-term in amplitudes between two sub-beams. 
	\\
	In contrast, Aharonov \textit{et al.} propose that, if a certain interaction with a general interaction strength $\alpha$ in a path generates detectable consequences, the property associated with the interaction is localised in that path. This is a statement about one localisation measurement in one path. When trying to make statements about all involved properties in all paths together, it needs to be assured that the interactions do not change the outcome of other localisation measurements. Therefore, the disturbance of the interactions needs to be small. This is achieved by choosing small interaction strengths $\alpha\ll1$ such that the interactions are weak. Due to the small disturbances, the localisations of the properties cannot be determined for a single neutron but only with the statistics of a large number of neutrons in an ensemble. If each weak interaction only effects a conspicuous consequence when applied in a different path, the interpretation of separated properties in an ensemble is proposed. In this interpretation, each property is sensitive to a different operation $\hat{O}$. The implementation of $\hat{O}$ then serves to locate the corresponding property and the quantification of the path occupation of a property is given by the operator's weak value \cite{Aharanov1988weakvalues, Duck89, Watanabe1955, Aharanov1964,Aharanov2001} written as
	\begin{equation}
		\label{WeakValue}
		\braket{\hat{O}}_{\mathrm{w}}=\frac{\bra{\mathrm{f}}\hat{O}\ket{\mathrm{i}}}{\braket{\mathrm{f}|\mathrm{i}}} \in\mathbb{C},
	\end{equation} 
	with the preselected initial state $\ket{\mathrm{i}}$ and the postselected final state $\ket{\mathrm{f}}$. 
	\\
	We ask the benevolent reader to regard all following formulations which touch on the apparent separation of properties as neutral between both possible interpretations. We will state our own viewpoint in the discussion.  
	\\
	\par 
	The first experimental implementation of a qCC was carried out by Denkmayr \textit{et al.} \cite{Denkmayr2014} in a two-path neutron interferometer with the properties of particle and spin. While a particle is supposed to be affected by an absorber, a spin interacts with magnetic fields. Implementing these interactions affected the detected mean intensity in one path and the interference contrast in the other path, giving rise to the perception of a spatial separation of particle and spin. 
	\\
	A selection of other implementations of the Cheshire Cat effect and some discussions can be found in \cite{Atherton15, Ashby2016, Correa2015, Stuckey2016, DUPREY20181, Das2019, Liu2020, Kim2021, Sau2022, Li2023}. Possible improvements of the experiment were suggested, such as the simultaneous realisation of all weak measurements  \cite{Aharonov_2013}, further consideration of not only the first-order but the second-order  consequences of the midway interaction \cite{Stuckey2016}, and the implementation of additional degrees of freedom as pointer systems \cite{DUPREY20181}. Critique was expressed, for instance, questioning whether the observed effect is purely quantum mechanical \cite{Atherton15} and concerning the midway interaction strength \cite{Kim2021}. 
	\\
	A generalised form of a qCC was proposed by Pan \cite{Pan2020} with arbitrarily many degrees of freedom and, in consequence, properties. Here, we report on the realisation of a three-path quantum Cheshire Cat in a neutron interferometric experiment by additionally choosing the energy as third property. 
	\begin{figure}
		\includegraphics[scale=0.8]{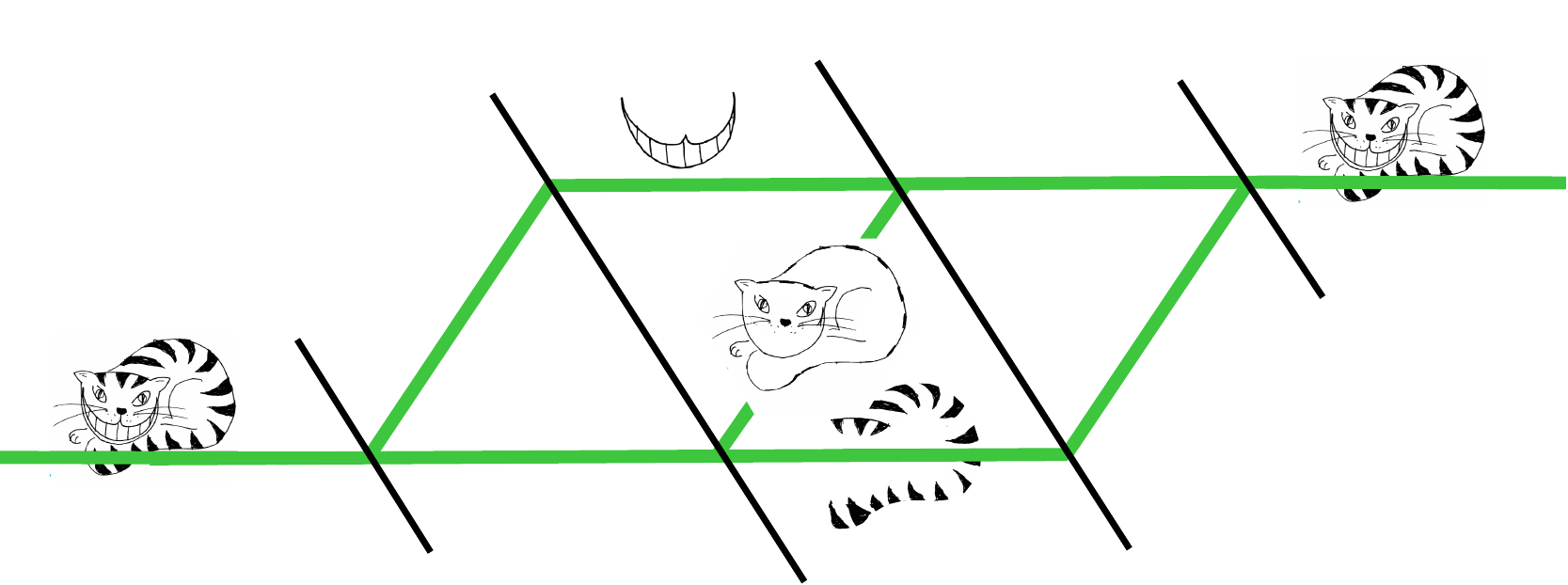}
		\caption{\textbf{Schematic of the paradoxical effect of the three-path quantum Cheshire Cat.} The cat is separated into three different parts inside the interferometer. This is in analogy to how the neutrons and their properties behave in the present experiment. The parts of the cat correspond to the neutron properties spin (grin), particle (body) and energy (stripes). The consequences to the interactions applied during the experiment may lead to the perception that the properties of the neutron are separated. 
		}
		\label{scheme}
	\end{figure}
	The schematic of the three-path qCC is illustrated in Fig.\,\ref{scheme}. Each part of the cat corresponds to a property of the neutron. When directly attributing the localisation of properties to consequences to local manipulations, the associations are as follows: a direct current (DC) spin-rotation affects the spin, a radio frequency (RF) rotation the energy and absorption the particle. The consequences to the weak interactions are observed and weak values are determined for quantification. By means of this extended version we seek to demonstrate how the qCC emerges from the geometrical relation between preselection, postselection, and the weak interactions which will make the essence of the qCC evident. 
	\\

	\section{Results}
	\label{results}
	
\begin{figure}
		\includegraphics[scale=0.6]{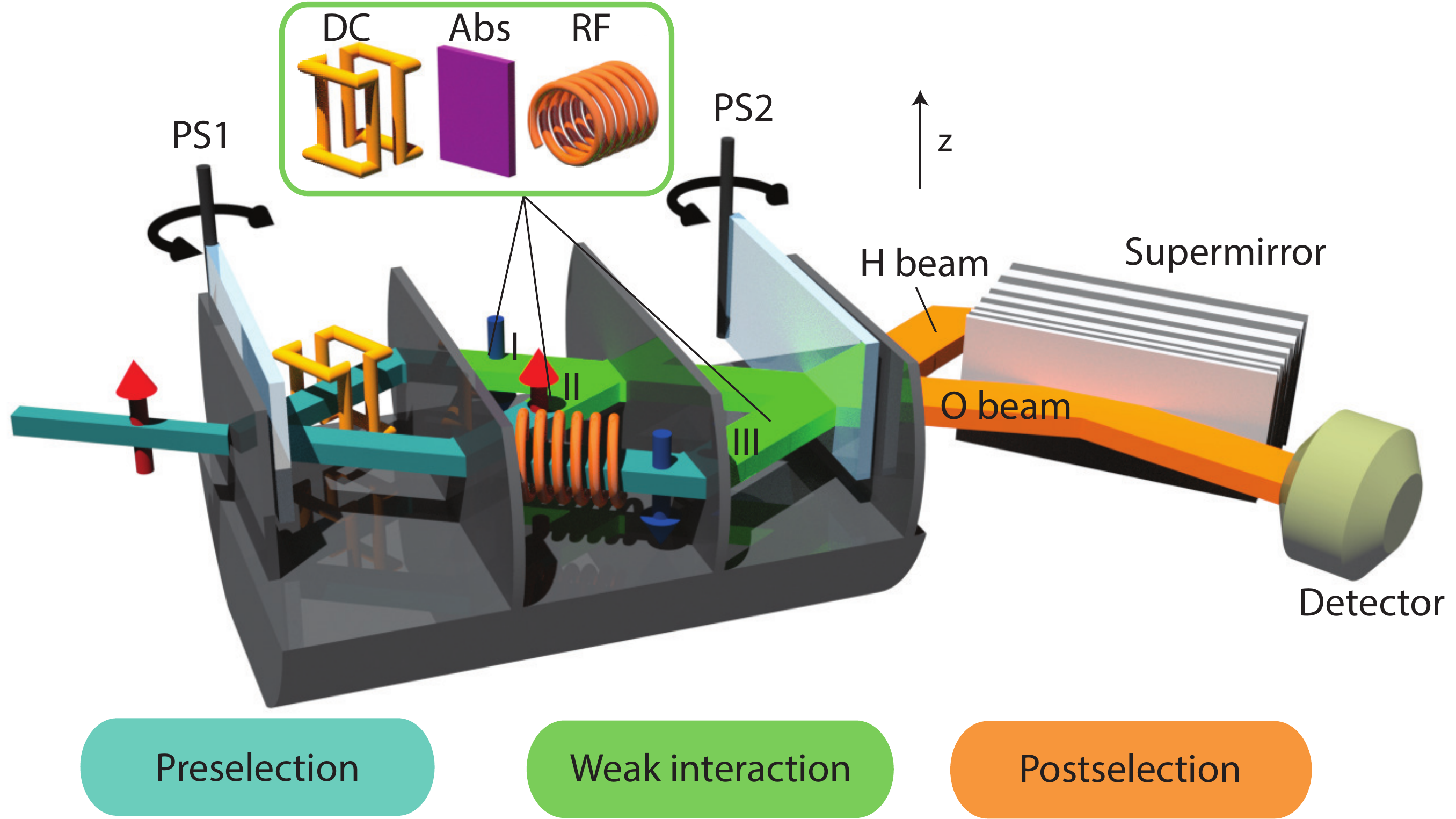}	
		\caption{\textbf{Setup of the neutron interferometer experiment downstream of monochromator and polariser.} An incoming neutron beam which is polarised in +z direction (red arrows) is split into three paths inside a perfect crystal interferometer. All sub-beams are recombined and the neutrons in the outgoing O beam are detected. The H beam is only used for monitoring. The experiment consists of three stages: first, the preselection or preparation stage (turquoise) where direct current (DC) and radio frequency (RF) spin manipulators flip the local spin vectors to the downward orientation (blue arrows) and produce three pairwise orthogonal sub-beams, cf.\ text and Eq.\,(\ref{preselection}). Second, the weak interaction stage (green) where one of three interactions, i.e.\ beam attenuation/absorption (abs) as well as DC and RF spin rotations, can be applied weakly in one of the three paths. Finally, the analysis or postselection (orange) where the phase shifters (PS) 1 and 2 determine the phases $\chi_1$, $\chi_2$, cf.\ Eq.\,(\ref{postselection}). At the recombination of the sub-beams and at the supermirror, respectively, the postselection projects the incoming state onto a specific phase relation between the sub-beams in the O beam and onto the up-spin state. 
		}
		\label{setup}
	\end{figure}

	\emph{Scheme and Theory.---}%
	\\
	The experiment was carried out at the neutron interferometry station S18 at the high-flux reactor of the Institut Laue-Langevin (ILL) in Grenoble, France. The neutrons are monochromatised with a perfect silicon crystal to a wavelength $\lambda=1.9$\,\AA\ and then polarised by a magnetic prism \cite{Badurek00FieldPrism} to the upward +z direction which defines the quantisation axis. We will use the symbols $\uparrow$ and $\downarrow$ in the notation to refer to up and down spin states, respectively. The arrows correspond to the +z and -z directions of the polarisation vector $\vec{P}=\braket{\psi|\vec{\hat{\sigma}}|\psi}$ where $\vec{\hat{\sigma}}=(\hat{\sigma}_\mathrm{x},\hat{\sigma}_\mathrm{y},\hat{\sigma}_\mathrm{z})^\mathrm{T}$ with the Pauli matrices $\hat{\sigma}_k$. The setup downstream of monochromator and polariser is depicted in Fig.\,\ref{setup}. The beam is split by the first two of four plates of a silicon perfect crystal interferometer into the three separated sub-beams indexed as $j\in\{\mathrm{I,II,III}\}$. After recombination of all three sub-beams, two outgoing beams are produced: the O beam in forward direction and the H beam in diffracted direction. The H beam is only used for monitoring. A spin analysis is implemented in the O beam by a polarising CoTi multilayer array, henceforth referred to as a supermirror. The intensities of O and H beam are recorded by $^3$He counting tubes. Inside the interferometer, two phase shifters (PS1 and PS2) control the phase relations between the three paths. Furthermore, if necessary, a weak spin or energy manipulation or a weak beam attenuation 
	is applied in the interferometer. Please note that all spin manipulators, although not constantly in use, are in place throughout the time of the experiment. 
	\\
	The experimental procedure can be divided into the three stages of preselection, weak interactions and postselection. The preselection is realised by monochromator, polarising magnetic prism, the beam splitters of the interferometer and two spin flippers in paths I and III. The spin flipper in path I induces a static DC spin flip and the one in path III an RF spin flip with a frequency $f=60$\,kHz of the oscillating magnetic field (see section \ref{DetailedAdjustment} for details of adjustment). The RF spin flip also changes the energy by $\Delta E=hf\approx 0.25$\,neV, shifting the initial kinetic energy $E_0\approx25$\,meV of the thermal neutrons to the new energy $E'=E_0-\Delta E$, with $\Delta E/E_0\approx10^{-8}$. The combined effect of aforementioned neutron optical components makes the separated sub-beams pairwise orthogonal when being recombined: while the spin orientation is up in path II and down in paths I and III, the latter two are in different energy states. Although the two energy states exhibit time-dependent interference on the micro-second scale, the detected counts are time-integrated such that no interference is observable and the two energy states, too, are orthogonal to each other \cite{Badurek86DoubleRes}. The two occupied energy levels and their respective two-level system behave like a pseudeospin system as applied earlier \cite{Sponar2008, Hasegawa2010, Shen2020, Lu2020}. We will therefore use the braket notation as an abbreviation to refer to the energy vectors as well as the path and spin vectors. The according triply entangled preselected initial state $\ket{\mathrm{i}}$ is then written as
	\begin{equation}
		\label{preselection}
		\ket{\mathrm{i}}=\frac{1}{\sqrt{3}}\bigg(\ket{\mathrm{I, \downarrow, E_0}}+\ket{\mathrm{II, \uparrow, E_0}}+\ket{\mathrm{III, \downarrow, E'}}\bigg).
	\end{equation} 
	\\
	Therein, all states from different Hilbert spaces associated with a sub-beam are written together in a single ket for each path. The postselection consists of PS1 and PS2 with their induced relative phases $\chi_1$ and $\chi_2$, the analysing crystal plates and the supermirror in the O-beam. The postselected final state $\ket{\mathrm{f}}$ is given by 
	\begin{equation}
		\label{postselection}
		\ket{\mathrm{f}}=\ket{\mathrm{f(\chi_1,\chi_2)}}= \frac{1}{\sqrt{3}} \ket{\uparrow} \left(\mathrm{e}^{\mathrm{i}(\chi_2-\chi_1)}\ket{\mathrm{I}}+\mathrm{e}^{\mathrm{i}(\chi_1+\chi_2)}\ket{\mathrm{II}}+\mathrm{e}^{\mathrm{i}(\chi_1-\chi_2)}\ket{\mathrm{III}}\right). 
	\end{equation} 
	We choose to attribute both phase shifts to the postselection rather than the preparation; both approaches are equivalent. The overlap $\braket{\mathrm{f|i}}$ between pre- and postselection has only a single non-zero term, coming from path II, while the components from the other paths in the initial state $\ket{\mathrm{i}}$ are orthogonal to $\ket{\mathrm{f}}$ such that their contributions to the overlap are zero. This means that, given the preselection, only the component of the sub-beam through path II is postselected. We will therefore refer to path II as the reference beam in our experiment. The other paths I and III can contribute to the postselected state, however, when additional weak interactions are applied as described in the next paragraph. In contrast to the generalised proposal by Pan \cite{Pan2020} (see section \textit{Discussion}), our postselection is not energy selective. Nonetheless, our setup exhibits the same effects in the limit of small interaction strengths as clarified in the section \textit{Discussion}.
	\\
	In the weak interaction stage between pre- and postselection, we choose to apply a weak DC or RF spin rotation, or a beam attenuation. (For the reference measurements, all weak interactions are turned off.) The interaction strengths are tuned by the DC/RF spin rotation angles $\alpha_\mathrm{rot}=\pi/9\,\hat{=}\,20\,^\circ$, and the absorption coefficient $\mathcal{A}=0.1$ as realised by an Indium foil of 0.125\,mm thickness. The absorption differs from the cases of DC/RF spin manipulations as it is not a unitary operation; the conceptual consequences will be explained throughout the rest of the article. We apply only one interaction in one beam at a time. Any of the three interactions can be applied to any of the three paths, obtaining nine different situations. All interactions are weak and create only small disturbances on the initial state. By combining the results of each single situation, we deduce the localisations of properties of the detected neutrons between pre- and postselection. 
	\\
	The relevant Pauli matrices for the spin and energy flips in the DC and RF cases are given by 
	\begin{equation}
		\hat{\sigma}^\mathrm{DC}_\mathrm{x}=\ket{\uparrow}\bra{\downarrow}+\ket{\downarrow}\bra{\uparrow}=
		\begin{pmatrix}
			0 & 1\\
			1 & 0
		\end{pmatrix}
		_\mathrm{spin}
	\end{equation}
	and
	\begin{equation}
		\hat{\sigma}^\mathrm{RF}_\mathrm{x}=\hat{\sigma}^\mathrm{DC}_\mathrm{x}\otimes\bigg(\ket{E'}\bra{E_0}+\ket{E_0}\bra{E'}\bigg)=
		\begin{pmatrix}
			0 & 1\\
			1 & 0
		\end{pmatrix}
		_\mathrm{spin}\otimes
		\begin{pmatrix}
			0 & 1\\
			1 & 0
		\end{pmatrix}
		_\mathrm{energy},
	\end{equation}\\
	respectively. The path projectors $\hat{\Pi}_j=\ket{j}\bra{j}$ indicate in which path an operation or manipulation is conducted. Then the unitary operators for spin and energy rotations $\hat{U}^\mathrm{DC}_j$ and $\hat{U}^\mathrm{RF}_j$, with the rotation angle $\alpha_\mathrm{rot}$ around the x axis in path $j$, while leaving the states in the other paths unchanged, can be expressed as (detailed calculation in section \ref{DetailedCalculation})
	
	\begin{align} 
		\begin{split}
			\label{EulerRepresentation}
			\hat{U}^\mathrm{DC}_j(\alpha_\mathrm{rot})&=\mathrm{exp}\left(-\mathrm{i}\frac{\alpha_\mathrm{rot}}{2}\hat{\sigma}^\mathrm{DC}_\mathrm{x}\hat{\Pi}_j\right)\\
			&=\mathds{1}-\left(1-\cos\left(\frac{\alpha_\mathrm{rot}}{2}\right)\right)\hat{\Pi}_j
			-\mathrm{i}\sin\left(\frac{\alpha_\mathrm{rot}}{2}\right)\hat{\sigma}^\mathrm{DC}_\mathrm{x}\hat{\Pi}_j\ \ \mathrm{and}\\ 
			\hat{U}^\mathrm{RF}_j(\alpha_\mathrm{rot})&=\mathrm{exp}\left(-\mathrm{i}\frac{\alpha_\mathrm{rot}}{2}\hat{\sigma}^\mathrm{RF}_\mathrm{x}\hat{\Pi}_j\right)\\
			&=\mathds{1}-\left(1-\cos\left(\frac{\alpha_\mathrm{rot}}{2}\right)\right)\hat{\Pi}_j
			-\mathrm{i}\sin\left(\frac{\alpha_\mathrm{rot}}{2}\right)\hat{\sigma}^\mathrm{RF}_\mathrm{x}\hat{\Pi}_j.
		\end{split}
	\end{align} 
	The x direction is always defined by the beam direction in the respective section of the setup (see section \ref{DetailedAdjustment} for further explanation). Equation (\ref{EulerRepresentation}) indicates that the DC (RF) spin rotation reduces the amplitude of the original spin-component (spin/energy-component) in the corresponding path $j$ from 1 to $\cos(\alpha_\mathrm{rot}/2)$ and creates a spin-flipped (spin/energy-flipped) component of amplitude $-\mathrm{i}\sin(\alpha_\mathrm{rot}/2)$. In the limit of small $\alpha_\mathrm{rot}$, $\sin(\alpha_\mathrm{rot}/2)$ is linear in $\alpha_\mathrm{rot}/2$, while the change of the original component, $1-\cos(\alpha_\mathrm{rot}/2)$, is only a smaller one proportional to $\alpha_\mathrm{rot}^2/8$.
	\\
	The operator $\hat{A}^\mathrm{Abs}_j(\cal{A})$  for a weak absorption is written as
	\begin{equation}
		\label{absorptionOperator}
		\hat{A}^\mathrm{Abs}_j(\mathcal{A})=\mathds{1}-(1-\sqrt{1-\mathcal{A}})\hat{\Pi}_j. 
	\end{equation} 
	It simply describes an attenuation in path $j$ while all other paths are undisturbed. 
	\\
	In the following, we will show how the observed intensities in the experiment are connected with the weak values of the local spin and energy as well as path observables. For a description consistent with the notation of conventional weak values as in Eq.\,(\ref{WeakValue}), we introduce two ancillary states
	\begin{equation}
		\ket{\mathrm{f_0}}=\ket{\mathrm{f}}\ket{\mathrm{E_0}}\ \ \mathrm{and}\ \ \ \ket{\mathrm{f'}}=\ket{\mathrm{f}}\ket{\mathrm{E'}}
	\end{equation}
	and define the weak value for the hypothetical energy selection onto $\ket{\mathrm{E}_0}$ as 
	\begin{equation}
		\braket{\hat{O}}^\mathrm{E_0}_\mathrm{w}=\frac{\bra{\mathrm{f_0}}\hat{O}\ket{\mathrm{i}}}{\braket{\mathrm{f_0}|\mathrm{i}}}.
	\end{equation}
	As we will see shortly, our results will be described by the weak values of the operators $\hat{\sigma}^\mathrm{DC}_\mathrm{x}\hat{\Pi}_j$, $\hat{\Pi}_j$ and $\hat{\sigma}^\mathrm{RF}_\mathrm{x}\hat{\Pi}_j$ where $j$ denotes the path. The first operator represents the x component of the spin in path $j$, while the latter one is the x component of the energy observable in path $j$ which is associated with flips in the energy system. The calculation of their weak values is straightforward for the initial and final states given in Eqs.\,(\ref{preselection},\ref{postselection}) and yields
	\begin{align}
		\begin{split}
			\label{weakValueCalc}
			\braket{\hat\sigma^\mathrm{DC}_{\mathrm{x}}\hat{\Pi}_j}_\mathrm{w}^\mathrm{E_0}&=\delta_{j,\mathrm{I}} \mathrm{e}^{2\mathrm{i}\chi_1}, \\
			\braket{\hat{\Pi}_j}_\mathrm{w}^\mathrm{E_0}&=\delta_{j,\mathrm{II}},\ \ \ \ \mathrm{and}\\
			\braket{\hat\sigma^\mathrm{RF}_{\mathrm{x}}\hat{\Pi}_j}_\mathrm{w}^\mathrm{E_0}&=\delta_{j,\mathrm{III}} \mathrm{e}^{2\mathrm{i}\chi_2}.
		\end{split}
	\end{align} 
	Please note that the Kronecker deltas $\delta_{i,j}$ give the absolute value of the respective weak value. Quantifying the location of properties through the weak values, an absolute value of an operator's weak value of 1, which is one of the operator's eigenvalues, is attributed to finding the corresponding property in the considered path. An absolute value of the weak value of zero excludes finding the property in that path. 
	\\
	With these expressions, the intensity $I$ measured in the postselected output port of the interferometer, with a weak DC spin rotation applied (DC case) in path $j$, is written as (details in section \ref{DetailedCalculation}) 
	\begin{align}
		\begin{split}
			\label{CalcDC}
			I_{j}^{\mathrm{DC}}(\chi_1)
			=&\left| \braket{\mathrm{f}|\hat{U}^\mathrm{DC}_j(\alpha_\mathrm{rot})|\mathrm{i}}\right|^2\\
			=&\left| \braket{\mathrm{f|i}}\right| ^2\bigg[1+\alpha_\mathrm{rot}\mathrm{Im}\left\{ \braket{\hat{\sigma}^\mathrm{DC}_\mathrm{x}\hat{\Pi}_j}^\mathrm{E_0}_\mathrm{w}\right\}
			+\frac{\alpha_\mathrm{rot}^2}{4}\left(\frac{\left| \bra{\mathrm{f_0}}\hat{\sigma}^\mathrm{DC}_\mathrm{x}\hat{\Pi}_j
				\ket{\mathrm{i}}\right| ^2}{\left| \braket{\mathrm{f|i}}\right| ^2}+\frac{\left| \bra{\mathrm{f'}}\hat{\sigma}^\mathrm{DC}_\mathrm{x}\hat{\Pi}_j
				\ket{\mathrm{i}}\right| ^2}{\left| \braket{\mathrm{f|i}}\right| ^2}\right)\\
			&-\frac{\alpha_\mathrm{rot}^2}{4}\mathrm{Re}\left\{\braket{\hat{\Pi}_j}^\mathrm{E_0}_\mathrm{w} \right\} +\mathcal{O}(\alpha_\mathrm{rot}^3)\bigg]\\
			=&\left| \braket{\mathrm{f|i}} \right| ^2 \bigg[1+\alpha_\mathrm{rot}\delta_{j,\mathrm{I}}\sin(2\chi_1)+\frac{\alpha_\mathrm{rot}^2}{4} \left(\delta_{j,\mathrm{I}}-\delta_{j,\mathrm{II}}+\delta_{j,\mathrm{III}}\right)
			\bigg]
			+\mathcal{O}(\alpha_\mathrm{rot}^3).
		\end{split}
	\end{align} 
	The second and third lines describe the intensity in terms of weak values and terms closely resembling them. The last line gives the expected intensity directly in terms of phase shifter orientation and selected path number. An intensity oscillation with an amplitude of order $\alpha_\mathrm{rot}$ can emerge under the condition of the Kronecker delta $\delta_{j,\mathrm{I}}$, which still gives the absolute value of the respective weak value. The condition is met when the weak DC spin rotation is applied in path I, which inverts part of the prepared down spin state orthogonal to the reference state to the up spin state which is parallel. The down component is filtered out by the supermirror of the postselection and the up component is transmitted to the detector. Simultaneously to the emergence of the intensity oscillation, the mean intensity is increased by the additional parallel component with order $\alpha_\mathrm{rot}^2$ as also pointed out in \cite{Stuckey2016}. The same mean intensity increase is expected from a weak DC spin rotation in the RF-flipped path III because a similar spin-up component is created which is transmitted through the supermirror. No intensity oscillation due to the differing energies is observable, though, in our time-integrating detection mode. In contrast, by applying the weak DC spin rotation in path II, the portion of the reference beam accepted by the supermirror is decreased with order $\alpha_\mathrm{rot}^2$. For small $\alpha_\mathrm{rot}$, the first order term is dominant and the consequences of a weak DC spin rotation on the intensity in path I are conspicuous. 
	\\
	A similar result can be derived assuming a weak RF spin rotation (RF case) applied in \mbox{path $j$}:
	\begin{align}
		\begin{split}
			\label{CalcRF}
			I_{j}^{\mathrm{RF}}(\chi_2)
			=&\left| \braket{\mathrm{f}|\hat{U}^\mathrm{RF}_j(\alpha_\mathrm{rot})|\mathrm{i}}\right|^2\\
			=&\left| \braket{\mathrm{f|i}}\right| ^2\bigg[1+\alpha_\mathrm{rot}\mathrm{Im}\left\{ \braket{\hat{\sigma}^\mathrm{RF}_\mathrm{x}\hat{\Pi}_j}^\mathrm{E_0}_\mathrm{w}\right\}
			+\frac{\alpha_\mathrm{rot}^2}{4}\left(\frac{\left| \bra{\mathrm{f_0}}\hat{\sigma}^\mathrm{RF}_\mathrm{x}\hat{\Pi}_j
				\ket{\mathrm{i}}\right| ^2}{\left| \braket{\mathrm{f|i}}\right| ^2}+\frac{\left| \bra{\mathrm{f'}}\hat{\sigma}^\mathrm{RF}_\mathrm{x}\hat{\Pi}_j
				\ket{\mathrm{i}}\right| ^2}{\left| \braket{\mathrm{f|i}}\right| ^2}\right)\\
			&-\frac{\alpha_\mathrm{rot}^2}{4}\mathrm{Re}\left\{\braket{\hat{\Pi}_j}^\mathrm{E_0}_\mathrm{w} \right\} +\mathcal{O}(\alpha_\mathrm{rot}^3)\bigg]\\
			=&\left| \braket{\mathrm{f|i}} \right| ^2 \bigg[1+\alpha_\mathrm{rot}\delta_{j,\mathrm{III}}\sin(2\chi_2)+\frac{\alpha_\mathrm{rot}^2}{4} \left(\delta_{j,\mathrm{I}}-\delta_{j,\mathrm{II}} +\delta_{j,\mathrm{III}}\right)
			\bigg]+\mathcal{O}(\alpha_\mathrm{rot}^3)
		\end{split} 
	\end{align} 
	The results in the DC and RF cases are similar up to the exchange of $\delta_{j,\mathrm{I}}$, $\delta_{j,\mathrm{III}}$, and $\chi_1$ in the DC case, respectively, for $\delta_{j,\mathrm{III}}$, $\delta_{j,\mathrm{I}}$, and $\chi_2$ in the RF case.  
	\\
	In the third case of an added weak absorber (absorber case), the intensity is described as (details in section \ref{DetailedCalculation}) 
	\begin{align}
		\begin{split}
			\label{CalcAbs}
			I_{j}^{\mathrm{Abs}}
			=&\left| \braket{\mathrm{f}|\hat{A}^\mathrm{Abs}_j(\mathcal{A})|\mathrm{i}}\right|^2\\	
			=&\left| \braket{\mathrm{f|i}} \right| ^2\left[1-\mathcal{A}\braket{\hat{\Pi}_j}^\mathrm{E_0}_\mathrm{w}\right]\\
			=&\left| \braket{\mathrm{f|i}} \right| ^2 \left[1-\mathcal{A}\,\delta_{j,\mathrm{II}} \right]. 
		\end{split}
	\end{align} 
	
	In effect, only an attenuation of the sub-beam in path II with its prepared up-spin state will be registered after the postselection at the detector in the O beam. 
	\FloatBarrier
	
	\begin{figure}[ht]
		\centering
		\includegraphics[width=16cm]{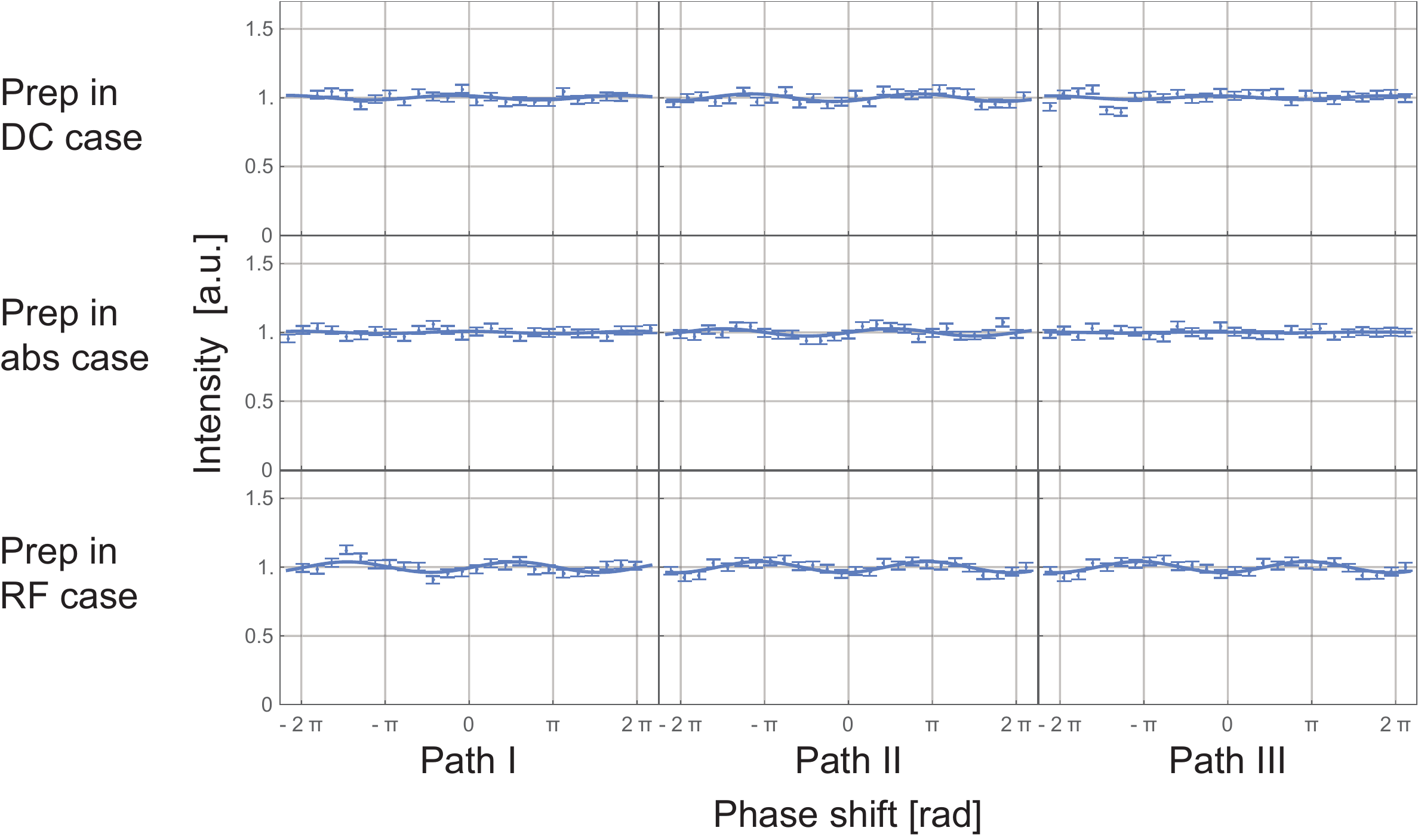}
		\caption{\textbf{Interferograms and their fits with preparation applied.} Intensities detected in O beam normalised by mean intensities plotted against phase shifts induced in the path specified at the bottom. Error bars indicate one standard deviation. The three rows of interferograms in the figure are obtained with different configurations for the preparation (prep) as written to the left which are used for different further measurements (see section \ref{DetailedAdjustment} for further explanation). The different configurations are used for further measurements. Contrasts are extracted from sinusoidal fits plotted as solid lines. The low contrasts $\leq 4\%$, given in Tab.\,\ref{contrastsPrep}, imply a good preparational quality, i.e.\ a high degree of orthogonality between the sub-beams. 
		}
		\label{IFGsprep} 
	\end{figure}
	
	\FloatBarrier
	
	\emph{Experimental Data. --- }
	
	\begin{table}
		\begin{tabular}{@{}lclclcl@{}}
			\multicolumn{7}{c}{\textbf{Contrasts with preparation}}\\
			\toprule
			\multirow{3}{3.5cm}[2pt]{\textbf{preparation for weak interaction}}&\phantom{abc}&\multicolumn{5}{c}{\textbf{phase shift in path}}\\
			\cmidrule{3-7}
			&&\makebox[3em]{I}&\phantom{a}&\makebox[3em]{II}&\phantom{a}&\makebox[3em]{III}	\\
			\hline
			\addlinespace
			DC &&1.7(9)\%&&1.7(9)\%&&1.3(11)\% \\ 
			\addlinespace
			Abs &&0.8(7)\%&&2.7(9)\%&&0.3(6)\%\\
			\addlinespace
			RF &&3.8(11)\%&&2.3(6)\%&&4.0(6)\%\\
			\addlinespace
			\bottomrule
		\end{tabular}
		\caption{Contrasts of fitted interferograms in Fig.\,\ref{IFGsprep} with preparations applied for the three different weak interactions but without the weak interactions themselves. The three rows are obtained with different configurations for the preparation as written to the left which are used for different further measurements (see section \ref{DetailedAdjustment}). The values in contrast quantify the quality of the preparation. \label{contrastsPrep}
		}
	\end{table}
	
\begin{figure}[ht]
		\centering
		\includegraphics[width=14cm]{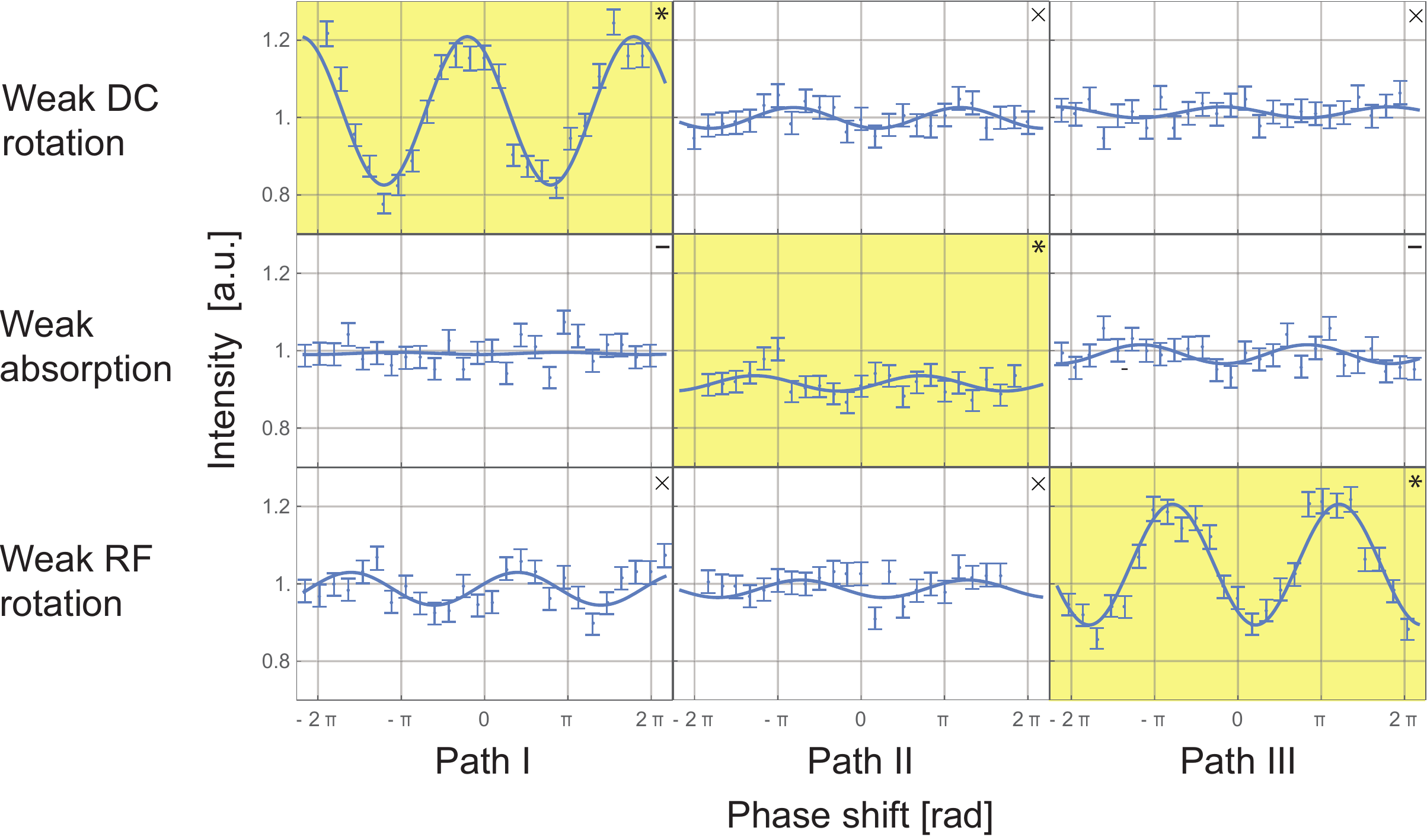}
		\caption{\textbf{Interferograms and their fits with preparation and weak interactions applied.} Intensities recorded in O beam normalised by mean intensities of corresponding preparational IFGs of Fig.\,\ref{IFGsprep} plotted against phase shifts induced in the path specified at the bottom. Error bars indicate one standard deviation, solid curves are fits. In each line, a different weak interaction is applied as labelled to the left. Most noticeable consequences compared to the preparational interferograms of Fig.\,\ref{IFGsprep} are found in coloured diagonal elements. Off-diagonal elements (shown with white background) exhibit only inconspicuous consequences. The symbols in the upper right corners indicate different kinds of situations described in the section \textit{Discussion}.
		}
		\label{IFGsWeak}
	\end{figure}

\begin{figure}[ht]
		\centering
		\includegraphics[width=8cm]{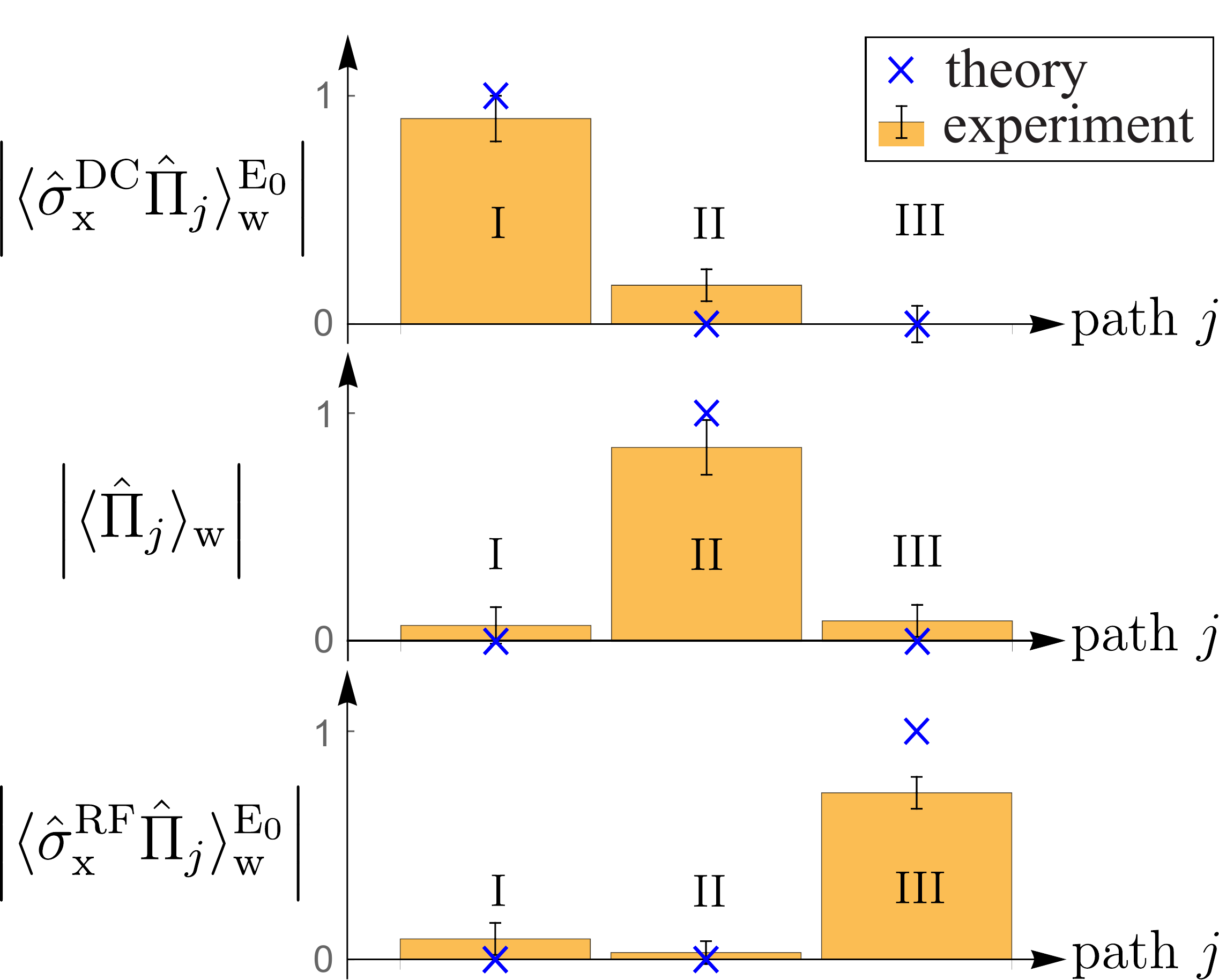}
		\caption{\textbf{Weak values presented graphically.} Absolute values of the relevant weak values extracted from interferograms in Figs.\,\ref{IFGsprep} and \ref{IFGsWeak} for each path $j$. For the path weak values of the absorber measurements, not the absolute value but the weak value itself is given. Blue crosses indicate the ideal theoretical absolute values of weak values which compose an identity matrix. 
		}
		\label{weakValuesGraph}
	\end{figure}

	The preparation is implemented by a DC flip in path I and an RF flip in path II. The measured interferograms (IFGs), depicted in Fig.\,\ref{IFGsprep}, will be called preparational IFGs. The fit function for all IFGs is of the form
	\begin{equation}
		\label{fitfunction}
		I(\chi)=I_0+B\sin(\omega\chi+\varphi), 
	\end{equation}
	with the mean intensity $I_0$ and an intensity oscillation with amplitude $B$, angular velocity $\omega$, the phase shifter orientation $\chi$, and the phase offset $\varphi$.  The preparational IFGs characterise the quality of orthogonality of the initial sub-states and are a reference for the quantitative data analysis. Phase shifts are implemented in all three paths, resulting in the three columns of Fig.\,\ref{IFGsprep}. In addition, three different experimental settings for the initial state, used for different further measurements as described in section \ref{DetailedAdjustment}, were realised, giving rise to the three rows in Fig.\,\ref{IFGsprep}. The obtained contrast values, specified in Tab.\,\ref{contrastsPrep}, quantify the quality of the preparation. We will refer to the following 3$\times$3 arrays of IFGs or numbers as matrices and to their diagonal, off-diagonal, and anti-diagonal elements as in a normal square matrix. 
	\\
	Finally, when separately applying one of the three weak interactions in one of the three sub-beams, the nine IFGs presented in Fig.\,\ref{IFGsWeak} were recorded which will be called weak-interaction IFGs. When comparing the weak-interaction IFGs of Fig.\,\ref{IFGsWeak} with the preparational IFGs of Fig.\,\ref{IFGsprep}, conspicuous consequences clearly appear in the coloured diagonal elements of Fig.\,\ref{IFGsWeak} where either significant intensity oscillations or a significant drop in count rate is produced. Weak values are extracted for all nine cases by comparing the measured IFGs of Figs.\,
	\ref{IFGsprep},\ref{IFGsWeak} with the predictions from Eqs.\,(\ref{CalcDC},\ref{CalcRF},\ref{CalcAbs}) in the limit of small interaction strengths. The detailed data analysis is given in section \ref{DetailedDataAnalysis}. The results are presented in Fig.\,\ref{weakValuesGraph} and Tab.\,\ref{weakValuesTable} and approximate the ideal identity matrix given by Eq.\,(\ref{weakValueCalc}). 
	
	\FloatBarrier
	
	\begin{table}
		\begin{tabular}{@{}cclclclcl@{}}
			\multicolumn{9}{c}{\textbf{Weak values}}\\
			\toprule
			&\phantom{abc}&\multicolumn{5}{c}{\textbf{path}}\\
			\cmidrule{3-7}
			\textbf{weak value}&&\makebox[3em]{I}&\phantom{a}&\makebox[3em]{II}&\phantom{a}&\makebox[3em]{III}&\phantom{aaa}&{$\sum_{\mathrm{I},\mathrm{II},\mathrm{III}}$}	\\
			\addlinespace[3pt]
			\hline
			\addlinespace[4pt]
			$\left|\braket{\hat\sigma^{\mathrm{DC}}_{\mathrm{x}}\hat\Pi_{j}}^\mathrm{E_0}_\mathrm{w}\right|$ &&0.90(10)&&0.17(7)&&0.00(8)&&1.07(15) \\
			\addlinespace[3.5pt]
			$\braket{\hat\Pi_j}^\mathrm{E_0}_\mathrm{w}$ &&0.07(8)&&0.85(12)&&0.09(7)&&1.01(16)\\
			\addlinespace
			$\left|\braket{\hat\sigma^{\mathrm{RF}}_{\mathrm{x}}\hat\Pi_{j}}^\mathrm{E_0}_\mathrm{w}\right|$ &&0.09(7)&&0.03(5)&&0.70(7)&&0.82(11)\\ 
			\addlinespace[9pt]
			\addlinespace[9pt]
			$\sum_{\mathrm{DC,Abs,RF}}$ &&1.06(15)&&1.05(15)&&0.79(13)&&-\\
			\addlinespace
			\bottomrule
		\end{tabular}
		\caption{Numerical presentation of the weak values of x-spin component, path operator, and energy transition operator of each path as presented graphically in Fig.\,\ref{weakValuesGraph}. While the absolute value of the weak values are extracted for the spin and energy observables, the path weak values are extracted directly. The sums of the weak values in each row and column are given at the bottom and right, respectively. 
		}
		\label{weakValuesTable}
	\end{table}

	\FloatBarrier
	\emph{Discussion.---}%
	\label{discussion}
	\\
	Let us compare our experiment with the generalized N-path qCC described by Pan \cite{Pan2020} which allows to pinpoint the emergence of the qCC mathematically. The generalised case considers N paths (indexed as $j$) and N-1 properties of two level systems (indexed as $p$). The two basis vectors in each Hilbert space of a property will be denoted as 1 and 0. The state vector entering the interferometer is assumed as $\ket{1,1,1,...}$. The sub-states in each path are prepared to be mutually orthogonal by flipping the respective state vector of property $p$ in path $j=p+1$. (Roman numbers indicating paths will henceforth appear in equations together with arabic numbers indicating properties.) The according preselection $\ket{\mathrm{i_N}}$ is denoted as 
	\begin{equation}
		\label{PreselectionAlok}
		\ket{\mathrm{i_N}}=\frac{1}{\sqrt{\mathrm{N}}}\bigg[ \ket{\mathrm{I}}\ket{1,1,1,...}+\ket{\mathrm{II}}\ket{0,1,1,...}+...+\ket{\mathrm{N}}\ket{1,...1,0}\bigg].
	\end{equation}
	
	The according postselection $\ket{\mathrm{f_N}}$ which is dependent on the phases $\chi_j$ of the phase shifters in path $j$ is chosen as 
	\begin{equation}
		\label{PostselectionAlok}
		\ket{\mathrm{f_N}}=\frac{1}{\sqrt{\mathrm{N}}}\left(\mathrm{e}^{\mathrm{i}\chi_1}\ket{\mathrm{I}}+\mathrm{e}^{\mathrm{i}\chi_2}\ket{\mathrm{II}}+...+\mathrm{e}^{\mathrm{i}\chi_\mathrm{N}}\ket{\mathrm{N}}\right) \ket{1,1,1,...}.
	\end{equation} 
	Please note that there is an analysis in the postselection for each property $p$ realised with a respective projection in the postselected state $\ket{\mathrm{f_N}}$. Similarly to the previous three-path consideration, the overlap $\mathrm{\braket{f_N|i_N}}$ between pre- and postselection has only a single non-zero term, coming from the component of path I, while the components from the other paths in the initial state $\ket{\mathrm{i_N}}$ are orthogonal to $\ket{\mathrm{f_N}}$ such that their contributions to the overlap are zero. This means that, given the preselection, only the component of the sub-beam through path I is postselected. We will therefore refer to path I as the reference beam and all others as non-reference beams in the generalised case. The operator for a manipulation of property $p$ in path $j$, while leaving the states in all other sub-beams unchanged, is given by
	\begin{align}
		\begin{split}
			\label{EulerRepresentationAlok}
			\hat{O}^p_j(\alpha)	&=\mathrm{exp}\left(-\mathrm{i}\frac{\alpha}{2}\hat{\sigma}^p_\mathrm{x}\hat{\Pi}_j\right)\\
			&=\mathds{1}-(1-\cos\frac{\alpha}{2})\hat{\Pi}_j
			-\mathrm{i}\sin\left(\frac{\alpha}{2}\right)\hat{\sigma}^p_\mathrm{x}\hat{\Pi}_j.
		\end{split}
	\end{align} 
	The weak values of the operators $\hat{\sigma}^p_\mathrm{x}\hat{\Pi}_j$ are written as
	\begin{equation}
		\braket{\hat{\sigma}^p_\mathrm{x}\hat{\Pi}_j}_\mathrm{w}=\delta_{j,p+1}\mathrm{e}^{\mathrm{i}(\chi_1-\chi_p)},
	\end{equation} 
	and the path weak values are written as
	\begin{equation}
		\braket{\hat{\Pi}_j}_\mathrm{w}=\delta_{j,1}.
	\end{equation}
	In addition to the $N-1$ properties indexed as $p$, the ``zeroth'' property of the generalised case would be the particle behaviour in path I such that a beam attenuation only has linear consequences on the mean intensity in path I, which is in analogy to Eq.\,(\ref{CalcAbs}) of the three-path consideration.
	\\
	It follows in an exact calculation (details in section \ref{DetailedCalculation}), without regarding the limit of small $\alpha$, that the intensity behaves as 
	\begin{align}
		\begin{split}
			\label{CalcAlok}
			I_j^p (\alpha)
			=&\left| \bra{\mathrm{f_N}}\hat{O}^p_j(\alpha)\mathrm{\ket{i_N}}\right|^2\\	
			=&\frac{1}{\mathrm{N^2}}\bigg[ 1+2\sin\left(\frac{\alpha}{2}\right)\mathrm{Im}\left\{\braket{\hat{\sigma}^p_\mathrm{x}\hat{\Pi}_j}_\mathrm{w}\right\}+\sin^2\left(\frac{\alpha}{2}\right)\left|\braket{\hat{\sigma}^p_\mathrm{x}\hat{\Pi}_j}_\mathrm{w}\right|^2-\sin^2\left(\frac{\alpha}{2}\right)\braket{\hat{\Pi}_j}_\mathrm{w}	\bigg]\\
			=&\left| \frac{1}{\mathrm{N}}\mathrm{e}^{-\mathrm{i}\chi_1}-\left(1-\cos\frac{\alpha}{2}\right)\frac{1}{\mathrm{N}}\mathrm{e}^{-\mathrm{i}\chi_1}\delta_{j,1}
			-\mathrm{i}\sin\left(\frac{\alpha}{2}\right)\frac{1}{\mathrm{N}}\mathrm{e}^{-\mathrm{i}\chi_p}\delta_{j,p+1}\right| ^2\\
			=&\frac{1}{\mathrm{N^2}}\bigg[ 1+2\delta_{j,p+1}\sin\left(\frac{\alpha}{2}\right)\sin(\chi_1-\chi_p)+\delta_{j,p+1}\sin^2\left(\frac{\alpha}{2}\right)-\delta_{j,1}\sin^2\left(\frac{\alpha}{2}\right)
			\bigg].	
		\end{split}
	\end{align} 
	This expresses the intensity through the related measures of weak values, amplitudes, and experimental parameters. The first line states that the intensity is determined by the geometrical relation between the initial state $\ket{\mathrm{i_N}}$, the unitary rotation implemented through $\hat{O}^p_j(\alpha)$, and the postselected state $\ket{\mathrm{f_N}}$. The second line gives the intensity in terms of weak values for given interaction strength $\alpha$. The weak values are multiplied with sine functions which depend on $\alpha$. It follows by expanding the intensity for small $\alpha$ that the weak values appear in every order of $\alpha$. Even though weak values were introduced as low order approximations \cite{Aharanov1988weakvalues}, they can be regarded as the expansion coefficients in the Taylor series \cite{Cheon2013, Dziewior2016} and can be used to describe the intensity for arbitrary interaction strengths $\alpha$. 
	In the third line, the intensity is expressed as the absolute squared of amplitudes from different paths; the first term is the amplitude from the reference state which is reduced by the second term if the condition $\delta_{j,1}=1$, or $j=1$, is met. This means any interaction implemented in the reference beam will reduce its postselected component through the projection in the first line. The third term in the third line is the amplitude of a non-reference beam with is produced if $\delta_{j,p+1}=1$, or $j=p+1$. In turn, the intensity then has a cross-term between the non-reference beam $j$ and the reference beam dependent on the locally induced phases $\chi_1, \chi_p$. 
	The last line gives the intensity dependent on the experimental parameters of the interaction strength $\alpha$ and the phases $\chi_1, \chi_p$. The intensity oscillation proportional to $\sin(\chi_1-\chi_p)$ is a consequence of the cross-term between the first and third term in the second line. The third and fourth terms in the last line are mean intensity changes which are conditioned through the Kronecker deltas $\delta_{j,p+1}$ and $\delta_{j,1}$. The data is analysed for second-order intensity changes in section \ref{ExperimentalResources}.
	\\
	We will go into detail now regarding the first line of Eq.\,(\ref{CalcAlok}) where the intensity is obtained by considering a rotation of the initial state $\ket{\mathrm{i}}$ and a projection on the final state $\ket{\mathrm{f_N}}$. Therefore, the geometrical relation in Hilbert space between the vectors of the postselected state and the intermediate state before postselection is essential. Any changes in the intensity compared to the preparational IFGs are a consequence of a weak interaction. As the weak interactions are unitary and the calculated intensity involves the projection to $\ket{\mathrm{f_N}}$, the consequences are expressed by sinusoidal functions in the last line of Eq.\,(\ref{CalcAlok}). By considering the geometrical relations, we can identify three different kinds of situations: 
	\\
	The first kind of situation arises when a weak interaction is applied to a non-reference beam. Let us consider a perturbation rotating the sub-state of a non-reference beam and thereby generating a state component that is parallel to the postselected state. This is equivalent to inverting a fraction of the sub-state from the orthogonal to the parallel component. Due to the behaviour given in Eqs.\,(\ref{EulerRepresentation}, \ref{EulerRepresentationAlok}), in the limit of $\alpha$ becoming zero, the magnitude of the following consequence on the intensity is linear to the interaction strength $\alpha$. The large consequence is identified with the behaviour proportional to $2\sin(\alpha/2)$ given in Eq.\,(\ref{CalcAlok}) for the exact calculation and, in the limit of $\alpha_\mathrm{rot}$ becoming zero, with the term proportional to $\alpha_\mathrm{rot}$ in Eqs.\,(\ref{CalcDC},\ref{CalcRF}). In these situations, the detected intensity is sensitive with respect to the weak interaction applied. At the same time, the parallel component causes an increased intensity proportional to $(+)\sin^2(\alpha/2)$ in Eq.\,(\ref{CalcAlok}) and proportional to $(+)\alpha^2_\mathrm{rot}$ in Eqs.\,(\ref{CalcDC},\ref{CalcRF}). 
	\\
	The second kind of situation arises if any rotation is applied to the reference beam. Then the amplitude of the postselected component is reduced. However, in comparison to the first kind of situation it is only a small consequence proportional to $(-)\sin^2(\alpha/2)$ in Eq.\,(\ref{CalcAlok}) and proportional to $(-)\alpha^2_\mathrm{rot}$ in Eqs.\,(\ref{CalcDC},\ref{CalcRF}). The intensity in these situations is robust with respect to the weak interaction. 
	\\
	The third kind of situation concerns the states of the non-reference beams again, now in combination with unitary rotations which do not produce a postselected component. Any consequence on the intensity is excluded by the Kronecker deltas and we conclude that in these situations the intensity is indifferent to the unitary rotations. 
	\\
	To first order in the weak interaction strength, the intensity dependences of the generalised and our experimental case are the same. The large first order consequences are seen in the diagonal elements in the weak interaction IFGs of Fig.\,\ref{IFGsWeak}, which are marked with asterisks (*) in the upper right corners of the graphs. In the same figure, second order consequences of the reduction in the mean intensities compared to the preparational IFGs are expected to appear in the upper and lower IFGs of the middle column. The small consequences are due to the robustness of the reference beam with respect to rotations which is indicated with crosses ($\times$). The differences between the general case and our three-path case can be seen in the corner elements of the anti-diagonal in Fig.\,\ref{IFGsWeak}: in the generalised case, these elements should behave indifferently. However, without energy projection in our postselection, we expect an increase in intensity proportional to $(+)\alpha^2_\mathrm{rot}$. This is caused by the up-spin components created by the weak interactions that produce additional counts without time-independent interference (explained in second paragraph of section \ref{results}). Additionally, the left and right IFGs in the absorber case in Fig.\,\ref{IFGsWeak} are indifferent in accordance to Eq.\,(\ref{CalcAbs}), as indicated with dashes (--). These situations do not involve rotations, however, and concern the location of the particle which is not explicitly regarded as a property by Pan. 
	\\
	\par
	Finally, we consider the weak values again. According to the geometrical considerations of the state vectors and rotations, if the absolute value of a weak value is zero, the intensity doesn't have a linear dependence to the interaction strength $\alpha$. Then, the intensity is either robust or indifferent to the weak interaction in the considered path. If the absolute value of a weak value is 1, it identifies a combination of path and weak interaction in which the intensity is sensitive to the weak interaction. 
	\\
	The sensitive behaviour can be interpreted as an interference effect emerging through a cross-term of amplitudes between two sub-beams, cf.\ Eq.\,(\ref{CalcAlok}). The magnitude of the cross-term is linear in $\alpha$ for small interaction strengths and therefore has conspicuous consequences on the intensity. Because the cross-term involves two paths, it offers an interpretation of delocalised properties in the interferometer. 
	\\
	The alternative interpretation proposed by Aharonov \textit{et al}.\ \cite{Aharonov_2013} quantifies the location of a property in a path through the weak values. A weak value of 1 is attributed to finding the property in that path; a value of zero excludes finding the property in that path. We identify these values with the absolute value of the weak values in the present experiment which is equivalent for phase shifter positions $\chi_1=\chi_2=0$. According to the latter interpretation, with the present results of Fig.\,\ref{weakValuesGraph} and Tab.\,\ref{weakValuesTable}, the neutron's x spin component is in path I, the particle in path II, and its x energy component in path III where the latter one is associated with an energy transition; a spatial separation of the neutron's properties inside the interferometer is observed.
	\\
	But how is the interpretation of separated properties compatible with the preselected state $\ket{\mathrm{i}}$ of Eq.\,(\ref{preselection}) where a specific value for spin and energy is attributed to each sub-state? 
	In some sense a particular weak value only gives partial information about an observable. 
	This is demonstrated by considering the expectation values of the path operators for the initial state of Eq.\,(\ref{preselection}) in the present experiment. At this point, the state of the neutron is distributed equally over all three paths, indicated by the expectation value $\braket{\mathrm{i}|\hat\Pi_j|\mathrm{i}} = 1/3$ for all paths. While we have so far considered only one particular final state $\ket{\mathrm{f}}$, given by Eq.\,(\ref{postselection}), one could in principle also monitor all relevant final states $\ket{\mathrm{f}_m}$, which are given in our setup by all combinations of the two z spin components 
	and the three spatial interferometer output ports. Besides O and H beam, the third output port is the side exit (not depicted in Fig.\,\ref{setup}) of the front interference loop between paths I and II. The order of how to index the output ports as $m$ is arbitrary. The set of output states $\{\ket{\mathrm{f_m}}\}$ is orthonormal and complete. Then the relation \cite{Hosoya2010,Hall2016}
	\begin{align}
		\begin{split}
			\braket{\mathrm{i}|\hat\Pi_j|\mathrm{i}}    &=\sum_m \braket{\mathrm{i}|\mathrm{f}_m} \braket{\mathrm{f}_m|\hat\Pi_j|\mathrm{i}}\\
			&=\sum_m p_m \braket{\mathrm{f}_m|\hat\Pi_j|\mathrm{i}}/\braket{\mathrm{f}_m|\mathrm{i}}\\
			&=\sum_m p_m \braket{\hat\Pi_j}_\mathrm{w}^m
		\end{split}
	\end{align}
	
	holds, where $p_m$ denotes the probability for given $\ket{\mathrm{i}}$ of reaching the final state $\ket{\mathrm{f}_m}$ and the path weak value $\braket{\hat\Pi_j}_\mathrm{w}^m$ is obtained for the final state $m$. Therefore, if we do not observe any intensity change when applying a weak beam attenuation in path I, it doesn't exclude a non-zero component to the state vector in that path. But it means that the component only contributes to an intensity in other exit channels given through other postselected states $\ket{\mathrm{f}_m}\neq\ket{\mathrm{f}}$. 
	However, for all neutrons that did reach our final state $\ket{\mathrm{f}}$ in the limit of zero weak interaction strength, we can in retrospect say that these neutrons never have been in \mbox{path I}. The two considered interpretations for the qCC are a separation of properties \cite{Aharonov_2013} on the one hand and an interference effect through the cross-term of amplitudes on the other hand. Only for the weak beam attenuations, both these interpretations agree that the weak values give the localisations in the interferometer of the neutrons found in our output port $\ket{\mathrm{f}}$. 
	\\
	As concerns the spin degree of freedom (likewise for the energy), the expectation value of the joint operator $\hat\sigma^\mathrm{DC}_\mathrm{x} \hat\Pi_j$ yields the $x$ component of the spin in path $j$. For our initial state, this value becomes zero in all paths, $\braket{\mathrm{i}|\hat\sigma^\mathrm{DC}_\mathrm{x} \hat\Pi_j|\mathrm{i}} = 0$, because the spin in each path is prepared in $\pm z$ direction and therefore has equal probabilities in $\pm x$ direction \cite{Stuckey2016}. Nevertheless, the weak value with our postselected final state $\braket{\hat\sigma^\mathrm{DC}_{\mathrm{x}}\hat{\Pi}_{\mathrm{I}}}_\mathrm{w}=\mathrm{e}^{2\mathrm{i}\chi_1}$ does not become zero, cf.\ Eq.\,(\ref{weakValueCalc}). The expectation value of zero results from the compensation 
	by a similar weak value with opposite sign in another output port of the interferometer, which in our setup is the down spin component of the side exit of the front loop
	. The opposite sign results form the phase shift of $\pi$ which always appears between the two output ports of an interferometer loop. 
	\\
	To conclude, all weak interactions applied in our experiment have similar consequences locally (in the respective path) but it is only the geometrical relation of the weak interaction to the pre- and postselection which makes its consequences visible or invisible in a particular final state. Presently, no alternative explanation -- such as a separation of properties -- is required to describe all observed phenomena. We suggest to regard the weak values as effective consequences in context of the qCC as opposed to the localisations of any form or component of spin and energy. While we cannot decide between the two considered interpretations with the present experiment, the unorthodox interpretation of separated properties would need extraordinary evidence to overcome the conventional description. Therefore, at the present moment, we do not attribute physical reality to the interpretation of separated properties, neither for an ensemble of nor for single neutrons themselves. 
	\\
	\par
	
	\FloatBarrier
	\section{Conclusion}
	%

	A three-path quantum Cheshire Cat is demonstrated with neutron interferometry; the neutron, its spin and its energy are found in different paths of the interferometer. In the experiment, a state preparation (preselection) as well as a state filtration (postselection) are implemented; even though the postselection is without energy discrimination, the effect of the quantum Cheshire Cat emerges as predicted by the theory. The conspicuous consequences of the local weak interactions are used to locate the properties of the neutron. Intensity oscillations are induced when weakly manipulating spin or energy to locate them, while an intensity reduction is induced for the weak beam attenuation applied when locating the neutron. These consequences are observed only for a particular interaction for each path. This suggests that the neutrons propagate through the interferometer, with the x spin component, particle, and energy taking different paths. A further analysis gives a more detailed explanation of the emergence of the consequences to the weak interactions. Only when a specific weak interaction is applied to a particular path, a certain component parallel to the reference state is generated; after the postselection, the final intensity contains the cross-term between this component and the reference. This suggests the possible explanation of the effect not as \textit{physical} but as \textit{effective} separation of properties in the interferometer.

	
	\FloatBarrier
	\section{Methods}
	
	\subsection{Adjustment Procedure}
	\label{DetailedAdjustment}
\begin{figure}[tb]
		\centering
		\includegraphics[width=18cm]{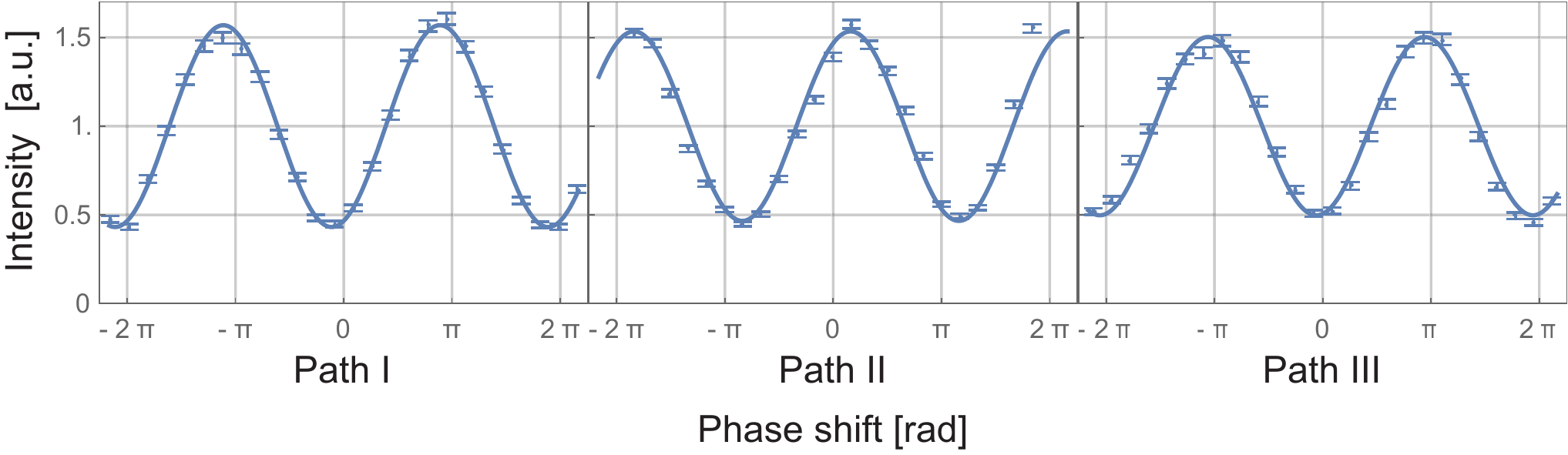}
		\caption{\textbf{Interferograms and their fits of empty interferometer.} Intensity detected in the O beam normalised by mean intensity plotted against phase shift induced in the path specified at the bottom. Error bars indicate one standard deviation. Contrasts are extracted from the respective sinusoidal fits plotted as solid lines. The contrasts from left to right of 57(1)\,\%, 53(3)\,\%, and 50(2)\,\% indicate the achievable level of coherence. 
		}
		\label{IFGsEmpty}
	\end{figure}

	We defined the z axis vertically and the x axis by the local beam direction in each section of the interferometer. Alternatively, one could differentiate between the orientations of the coils by explicitly defining global x and y axes. This would add a phase shift to certain spin components, as would a rigorous consideration of all neutron optical elements. But since our analysis of the data does not require a detailed justification of the phase shift between preparational and weak interaction IFGs, we simply determined the phase shift from the fitting parameters (see section \ref{DetailedDataAnalysis}).
	\\
	Each pair of sub-beams constitutes an interference loop. They are referred to as front, rear and outer loop which are composed, respectively, of beams I and II, beams II and III, and beams I and III. To confirm the initial coherence of the sub-beams in the interferometer, IFGs of the interferometer empty of any local fields or absorbers were recorded which are depicted in Fig.\,\ref{IFGsEmpty}. In the left case of Fig.\,\ref{IFGsEmpty} only PS1 was rotated and PS2 was oriented such that the rear loop passes on a maximum intensity in O direction (see Fig.\,\ref{setup}). Likewise, in the right case only PS2 was rotated and PS1 was oriented such that the front loop passes on a maximum intensity towards the last interferometer plate. In the middle case, both PS1 and PS2 are initially oriented such that a maximum intensity is acquired in O direction. The IFG is then recorded by simultaneously rotating both PS1 and PS2 to induce a relative phase between the reference beam II and the other two sub-beams. Contrasts $\geq$\,50\,\% were reached which indicate the moderate level of coherence achievable with our interferometer in the respective loops. The specific values, given in the caption of Fig.\,\ref{IFGsEmpty}, are used in the data analysis of section \ref{DetailedDataAnalysis} to compare the observed contrasts when applying weak interactions. 
	
\begin{figure}[ht]
		\centering
		\includegraphics[width=10cm]{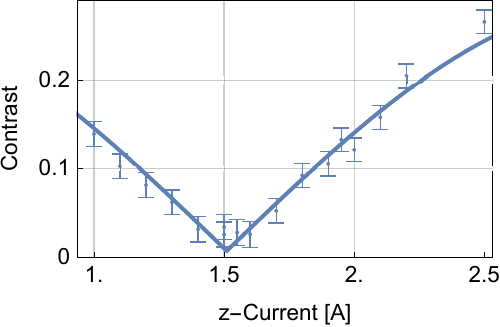}
		\caption{Contrasts recorded with an RF spin flipper in operation and the current $I$ for its local guide field amplification varied. Error bars indicate one standard deviation, solid curve is the fit. Around the minimum at $I_\mathrm{flip}$, the contrast has a local behaviour proportional to $|I-I_\mathrm{flip}|$ and therefore a sharper minimum than with a polarimetric approach via the intensity. }
		\label{interferometric adjustment}
	\end{figure}

	For both DC and RF coils, the currents for a spin flip have to be adjusted. These currents regulate the magnetic fields in x and z direction of each coil. The frequency for the RF coils was chosen to be 60\,kHz. This corresponds to a resonant field of about 20\,G=2\,mT local guide field strength. A global guide field of about 10 G was applied to allow for both RF and DC spin rotations: while this field was compensated to a net zero z field in coils when applying DC rotations, it was approximately doubled to meet the flip condition for RF rotations. This combination minimises inhomogeneities in the fields of the miniature spin rotators \cite{Geppert14, Danner19}. To roughly adjust and determine the flip currents in both the DC as well as RF coils, only the respective sub-beam was used, while all others were blocked by absorbers. This composes a polarimetric setup and the intensity was measured with varied z field and rotation currents. Estimates for guide field compensation (DC case) and strengthening (RF case) were determined as well as for the currents/amplitudes $I_\mathrm{flip}$ for the x fields of DC and RF flips. After that an interferometric method was used: as spin flips produce an orthogonal state compared to the reference beam of path II, interferograms have minimal contrast at the flip conditions. When recording several interferograms with slightly varied currents $I$ applied as described in the polarimetric case above, the resulting graph of contrasts has a sharper minimum due to the behaviour proportional to $|\sin(I-I_\mathrm{flip})|$ which is locally proportional to $I-I_\mathrm{flip}$ compared to the direct polarimetric approach with its cosine behaviour of the intensity which is locally proportional to $(I-I_\mathrm{flip})^2$ at its differentiable minimum. Since the weak interactions generate only sinusoidal intensity oscillations (in DC and RF case) with small amplitudes, it is especially important to ensure the lowest contrast possible from the preparation to begin with. This is assured with the interferometric approach. A typical adjustment scan is depicted in Fig.\,\ref{interferometric adjustment} where the contrast is found the lowest at 1.5\,A. To both sides of that value, the contrast is increasing at lowest order linearly. A residual contrast of about 3\% remains which is dominant in a small current interval at the minimum and which quantifies the overall spin manipulation efficiency of the setup. 
	\\
	\\
	To meet the resonance condition for spin rotations, the external guide field is locally suppressed by a compensation field for the weak DC rotations, while it is locally increased for the weak RF rotations. All local fields create stray fields and switching them on and off induces field offsets and inhomogeneities in the adjacent coils which lower the efficiency of their spin manipulations. The offsets can be compensated with our devices but the inhomogeneities cannot. 
	\\
	Therefore, in the weak RF measurements, we chose to leave the local guide field amplification permanently turned on and compensate the field offset in the adjacent coils. Then only the RF-field is turned on and off rather than both the RF and z fields. This technique lowers the efficiency of the manipulations through the inhomogeneities in the preparational cases but increases the overall efficiency when the weak interactions are applied. The technique cannot be applied in the DC case because it would create a zero-field region which induces depolarisation. Consequently, there are different preparational adjustments applied which ought to produce the same preselected state. This is the reason why there are multiple rows of preparational IFGs and their contrasts in Fig.\,\ref{IFGsprep} and Tab.\,\ref{contrastsPrep}.
	\\
	The weak interaction IFGs were recorded in combination with the preparational IFGs in an alternating ``on''/``off'' scheme, i.e.\ by turning the weak interaction on and off, before moving the phase shifter to the next orientation. This measurement protocol ensures the comparability of phase and contrast of the ``on'' and ``off'' IFGs which is needed in the data analysis. The IFGs with absorbers were not recorded in an `on''/``off'' scheme but right after each other while ensuring stable phase relations via the thermal control system.
	
	\subsection{Experimental Resources}
	\label{ExperimentalResources}
\begin{figure}[ht]
		\centering
		\includegraphics[width=10cm]{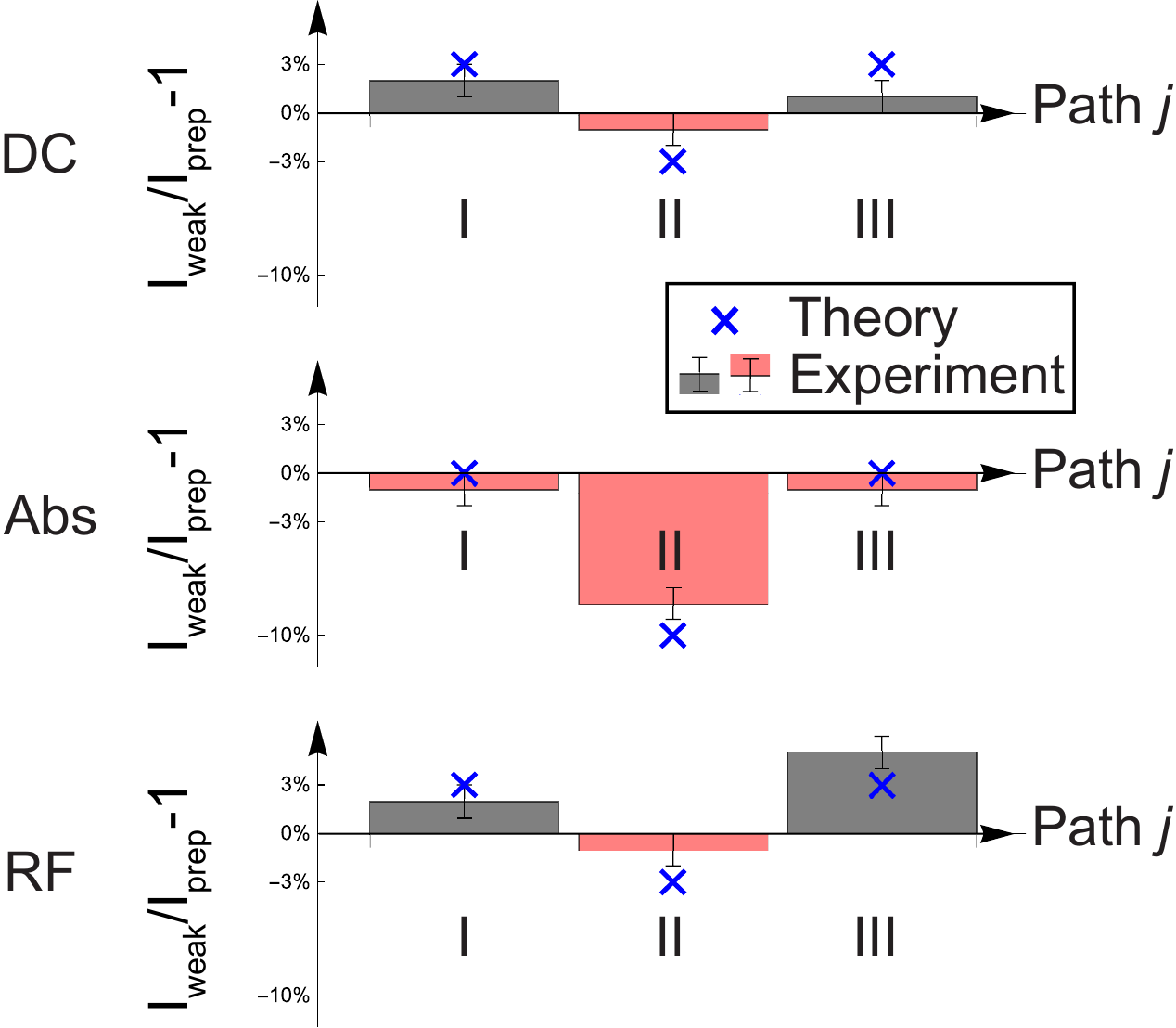}
		\caption{\textbf{Changes of the mean intensities presented graphically.} Graphical presentation of relative changes between mean intensities $I_\mathrm{weak}$ of weak measurement IFGs in Fig.\,\ref{IFGsWeak} normalised with mean intensities $I_\mathrm{prep}$ of preparational IFGs in Fig.\,\ref{IFGsprep} for each combination of weak interaction and path. Gray bars refer to an increase in intensity, pink ones to a decrease. Error bars indicate one standard deviation. Blue crosses indicate the values expected from theory. The normalised intensities are directly given in Tab.\,\ref{tableMeans}. In the absorber cases described by Eq.\,(\ref{CalcAbs}), an intensity drop proportional to $\mathcal{A}$ of 10\% is expected in path II. In the DC and RF cases described by Eqs.\,(\ref{CalcDC}, \ref{CalcRF}), intensity changes proportional to $\alpha_\mathrm{rot}^2$ of approximately $\pm3\%$ are expected. }
		\label{GraphMeans}
	\end{figure}

	For the given pre- and postselection of Eqs.\,(\ref{preselection}, \ref{postselection}), the contrast of IFGs is ideally zero. As can be seen from Fig.\,\ref{IFGsprep} and Tab.\,\ref{contrastsPrep}, lower contrasts were achieved in the DC and absorber cases compared to the RF case. This is expected when implementing the technique described two paragraphs above. On the other hand, the technique should increase the quality of the weak RF spin rotations. However, the weak value deviating the most from the theory is in the case of a weak RF interaction in path III which is obtained as $0.7$ compared to the prediction of 1, cf.\ Fig.\,\ref{weakValuesGraph}, Tab.\,\ref{weakValuesTable}, and Eq.\,(\ref{weakValueCalc}). As the coil for the preparational RF spin flip in path III is close to the coil inducing the weak RF rotation (see Fig.\,\ref{setup}), their interaction in terms of electrical oscillating circuits could have induced unintended additional spin manipulations. 
	\\
	Again, the weak value for the RF case in path III deviates from the theory. The deviations of all the other weak values from their expectations are of the magnitude of their errors. The errors are of the same magnitude for all elements as the decreased error of the amplitude for IFGs with low contrast is partly compensated by the increased error of the phase, see Eq.\,(\ref{errorPropagation}) in the data analysis below. Furthermore, we extract only the absolute value of the weak values for spin and energy observables. Thus, these off-diagonal weak values cannot be distributed symetrically around zero. 
	\\
	The energy changes produced with the RF coils are coupled with spin flips in our experiment. This is in principle avoidable when using a combination of an RF and a DC spin flipper instead. The first one flips the energy and spin vector, while the second one flips the spin vector back to the initial orientation. This effectively produces an energy change without spin manipulation. 
	\\
	Since both spin and energy are two-level systems, four possible combinations exist which are orthogonal to each other. The keen reader might have noticed that one of them, the state $\ket{\mathrm{\uparrow, E'}}$, was not mentioned, yet. This state has a particular character as it is expected to exhibit no conspicuous consequence to any \textit{single} weak interaction -- neither of DC nor RF spin rotations. The fourth state could only be produced by a combination of both a DC spin flip and a RF spin flip as described in the previous paragraph. The state does not imply a fourth property of the neutron, however, since the spin and energy systems are already manipulated by the other weak interactions used in our experiment. 
	\\
	When applying the weak interactions in our experiment, changes in the mean intensities compared to the preparational IFGs are expected in seven of the nine situations, where the term situation now refers to a combination of a specific weak interaction applied in a specific path. For the weak beam attenuations, the intensity changes directly give the path weak values of Eqs.\,(\ref{weakValueCalc}, \ref{CalcAbs}), Fig.\,\ref{weakValuesGraph}, and Tab.\,\ref{weakValuesTable}. For the unitary spin/energy manipulations, the mean intensity changes correspond to the terms proportional to $\pm\alpha_\mathrm{rot}^2$ in Eqs.\,(\ref{CalcDC}, \ref{CalcRF}). (In the exact calculation of Eq.\,(\ref{CalcAlok}), the mean intensity changes are represented by the terms proportional to $\pm\sin^2(\alpha/2)$.) In our experiment, the intensities are expected to increase by $\alpha^2_\mathrm{rot}/4\approx3\%$ when inducing weak unitary rotations in paths I or III, while a decrease of the same amount is expected for weak unitary rotations induced in path II. The measured intensity changes between the IFGs of Figs.\,\ref{IFGsprep} and \ref{IFGsWeak} are given in Fig.\,\ref{GraphMeans} and Tab.\,\ref{tableMeans}. The theoretical prediction and the experimental results show reasonable agreement. Their comparison suffices to establish higher order consequences which demonstrate that the intensity changes in all three paths through the unitary weak interactions as described in \cite{Denkmayr2014,Stuckey2016}. But a higher statistical precision will be necessary to quantitatively confirm the theoretically predicted intensity changes. 
	
	\FloatBarrier
	\begin{table}
		\begin{tabular}{@{}lclclcl@{}}
			\multicolumn{7}{c}{\textbf{Relative intensities}}\\
			\toprule
			\multirow{3}{2.3cm}[2pt]{\textbf{weak interaction}}&\phantom{abc}&\multicolumn{5}{c}{\textbf{path}}\\
			\cmidrule{3-7}
			&&\makebox[3em]{I}&\phantom{a}&\makebox[3em]{II}&\phantom{a}&\makebox[3em]{III}	\\
			\hline
			\addlinespace
			DC &&1.02(1)&&0.99(1)&&1.01(1) \\ 
			\addlinespace
			Abs &&0.99(1)&&0.92(1)&&0.99(1)\\
			\addlinespace
			RF &&1.02(1)&&0.99(1)&&1.05(1)\\ 
			\addlinespace
			\bottomrule
		\end{tabular}
		\caption{Numerical mean intensities of weak measurement IFGs in Figs.\,\ref{IFGsWeak} normalised with mean intensities of preparational IFGs in Fig.\,\ref{IFGsprep} for each combination of weak interaction and path as presented graphically as relative intensity changes in Fig.\,\ref{GraphMeans}. In the case of a weak absorber, solely a 10\% decrease is expected in path II, while the other paths are expected to be unaffected. In the DC and RF cases, consequences proportional to $\alpha_\mathrm{rot}^2$ are expected to change the mean intensities according to Eqs.\,(\ref{CalcDC},\ref{CalcRF}) by approximately $\pm3\%$.}
		\label{tableMeans}
	\end{table}
	\FloatBarrier
	
	\subsection{Detailed Calculations}
	\label{DetailedCalculation}
	
	The Euler representation of the unitary operators in Eqs.\,(\ref{EulerRepresentation},\ref{EulerRepresentationAlok}) is derived in the following.  
	\begin{align}
		\begin{split}
			\hat{O}^p_j(\alpha)	&=\mathrm{e}^{-\mathrm{i}\frac{\alpha}{2}\hat{\sigma}^p_\mathrm{x}\hat{\Pi}_j}\\
			&=\mathds{1}-\mathrm{i}\frac{\alpha}{2}\hat{\sigma}^p_\mathrm{x}\hat{\Pi}_j+\frac{1}{2!}(\mathrm{i}\frac{\alpha}{2}\hat{\sigma}^p_\mathrm{x}\hat{\Pi}_j)^2-\frac{1}{3!}(\mathrm{i}\frac{\alpha}{2}\hat{\sigma}^p_\mathrm{x}\hat{\Pi}_j)^3+\frac{1}{4!}(\mathrm{i}\frac{\alpha}{2}\hat{\sigma}^p_\mathrm{x}\hat{\Pi}_j)^4-...\\	
			&=\mathds{1}+\sum^\infty_{n=1}\frac{1}{(2n)!}\left(\mathrm{i}\frac{\alpha}{2}\hat{\sigma}^p_\mathrm{x}\hat{\Pi}_j\right)^{2n}-\sum^\infty_{n=0}\frac{1}{(2n+1)!}\left(\mathrm{i}\frac{\alpha}{2}\hat{\sigma}^p_\mathrm{x}\hat{\Pi}_j\right)^{2n+1}\\
			&=\mathds{1}+\sum^\infty_{n=1}\frac{(-1)^n}{(2n)!}\left(\frac{\alpha}{2}\right)^{2n}\hat{\Pi}_j-\mathrm{i}\sum^\infty_{n=0}\frac{(-1)^n}{(2n+1)!}\left(\frac{\alpha}{2}\right)^{2n+1}\hat{\sigma}^p_\mathrm{x}\hat{\Pi}_j\\
			&=\mathds{1}-\left(1-\cos\frac{\alpha}{2}\right)\hat{\Pi}_j-\mathrm{i}\sin\left(\frac{\alpha}{2}\right)\hat{\sigma}^p_\mathrm{x}\hat{\Pi}_j.
		\end{split}
	\end{align} 
	
	When a weak DC spin rotation is applied in path $j$ the measured intensity is given by Eq.\,(\ref{CalcDC}). Here we present the detailed derivation. We use the completeness relation 
	\begin{equation}
		\mathbb{1}=\ket{E_0}\bra{E_0}+\ket{E'}\bra{E'}, 
	\end{equation} 
	such that
	\begin{align}
		\begin{split}
			\label{calcDCdetailed}
			I_{j}^{\mathrm{DC}}(\chi_1)
			=&\left| \braket{\mathrm{f}|\hat{U}^\mathrm{DC}_j|\mathrm{i}}\right|^2
			=\left| \bra{\mathrm{f}} \mathrm{exp}\left(-\mathrm{i}\frac{\alpha_\mathrm{rot}}{2}\hat{\sigma}^\mathrm{DC}_\mathrm{x}\hat{\Pi}_j\right)\ket{\mathrm{i}}\right|^2\\	
			=&\left|\bra{\mathrm{f}}\left[\mathds{1}-\mathrm{i}\frac{\alpha_\mathrm{rot}}{2}\hat{\sigma}^\mathrm{DC}_\mathrm{x}\hat{\Pi}_j-\frac{\alpha_\mathrm{rot}^2}{8}\hat{\Pi}_j+\mathcal{O}(\alpha_\mathrm{rot}^{3})\right]\ket{\mathrm{i}}\right|^2\\
			=&\left| \braket{\mathrm{f|i}}
			-\mathrm{i}\frac{\alpha_\mathrm{rot}}{2}\bra{\mathrm{f}}\hat{\sigma}^\mathrm{DC}_\mathrm{x}\hat{\Pi}_j
			\ket{\mathrm{i}}
			-\frac{\alpha_\mathrm{rot}^2}{8}\bra{\mathrm{f}}\hat{\Pi}_j\ket{\mathrm{i}}
			+\mathcal{O}(\alpha_\mathrm{rot}^{3})\right| ^2\\
			=&\left| \braket{\mathrm{f|i}}\right| ^2
			+\alpha_\mathrm{rot}\mathrm{Im}\left\{ \braket{\mathrm{i|f}}\bra{\mathrm{f}}\hat{\sigma}^{\mathrm{DC}}_\mathrm{x}\hat{\Pi}_j	\ket{\mathrm{i}}\right\}
			+\frac{\alpha_\mathrm{rot}^2}{4}\left| \bra{\mathrm{f}}\hat{\sigma}_\mathrm{x}\hat{\Pi}_j
			\ket{\mathrm{i}}\right| ^2\\
			&-\frac{\alpha_\mathrm{rot}^2}{4}\mathrm{Re}\left\{ \braket{\mathrm{i}|\mathrm{f}}\bra{\mathrm{f}}\hat{\Pi}_j
			\ket{\mathrm{i}}\right\} +\mathcal{O}(\alpha_\mathrm{rot}^3)\\
			=&\left| \braket{\mathrm{f|i}}\right| ^2+\alpha_\mathrm{rot}\mathrm{Im}\left\{ \braket{\mathrm{i|f}}\left(\ket{\mathrm{E_0}}\bra{\mathrm{E_0}}+\ket{\mathrm{E'}}\bra{\mathrm{E'}}\right)\bra{\mathrm{f}}\hat{\sigma}^\mathrm{DC}_\mathrm{x}\hat{\Pi}_\mathrm{j}\ket{\mathrm{i}}\right\}\\
			&+\frac{\alpha_\mathrm{rot}^2}{4} \bra{\mathrm{i}}\hat{\sigma}^\mathrm{DC}_\mathrm{x}\hat{\Pi}_\mathrm{j}
			\ket{\mathrm{f}}\bra{\mathrm{f}}\hat{\sigma}^\mathrm{DC}_\mathrm{x}\hat{\Pi}_\mathrm{j}
			\ket{\mathrm{i}}-\frac{\alpha_\mathrm{rot}^2}{4}\mathrm{Re}\left\{\braket{\mathrm{i|f}}\left(\ket{\mathrm{E_0}}\bra{\mathrm{E_0}}+\ket{\mathrm{E'}}\bra{\mathrm{E'}}\right)\bra{\mathrm{f}}\hat{\Pi}_\mathrm{j}
			\ket{\mathrm{i}}\right\}+\mathcal{O}(\alpha_\mathrm{rot}^3)\\ 
			=&\left| \braket{\mathrm{f|i}}\right| ^2+\alpha_\mathrm{rot}\mathrm{Im}\left\{ \braket{\mathrm{i|f_0}}\bra{\mathrm{f_0}}\hat{\sigma}^\mathrm{DC}_\mathrm{x}\hat{\Pi}_\mathrm{j}\ket{\mathrm{i}}+\braket{\mathrm{i|f'}}\bra{\mathrm{f'}}\hat{\sigma}^\mathrm{DC}_\mathrm{x}\hat{\Pi}_\mathrm{j}\ket{\mathrm{i}}\right\}\\
			&+\frac{\alpha_\mathrm{rot}^2}{4} \bra{\mathrm{i}}\hat{\sigma}^\mathrm{DC}_\mathrm{x}\hat{\Pi}_\mathrm{j}
			\ket{\mathrm{f}}\left(\ket{\mathrm{E_0}}\bra{\mathrm{E_0}}+\ket{\mathrm{E'}}\bra{\mathrm{E'}}\right)\bra{\mathrm{f}}\hat{\sigma}^\mathrm{DC}_\mathrm{x}\hat{\Pi}_\mathrm{j}
			\ket{\mathrm{i}}
			-\frac{\alpha_\mathrm{rot}^2}{4}\mathrm{Re}\left\{\braket{\mathrm{i|f_0}}\bra{\mathrm{f_0}}\hat{\Pi}_\mathrm{j}
			\ket{\mathrm{i}}+\braket{\mathrm{i|f'}}\bra{\mathrm{f'}}\hat{\Pi}_\mathrm{j}
			\ket{\mathrm{i}}\right\}\\
			&+\mathcal{O}(\alpha_\mathrm{rot}^3)\\ 
			=&\left| \braket{\mathrm{f|i}}\right| ^2+\alpha_\mathrm{rot}\mathrm{Im}\left\{ \braket{\mathrm{i|f_0}}\bra{\mathrm{f_0}}\hat{\sigma}^\mathrm{DC}_\mathrm{x}\hat{\Pi}_\mathrm{j}\ket{\mathrm{i}}\right\}+\frac{\alpha_\mathrm{rot}^2}{4}\left(\left| \bra{\mathrm{f_0}}\hat{\sigma}^\mathrm{DC}_\mathrm{x}\hat{\Pi}_\mathrm{j}
			\ket{\mathrm{i}}\right| ^2+\left| \bra{\mathrm{f'}}\hat{\sigma}^\mathrm{DC}_\mathrm{x}\hat{\Pi}_\mathrm{j}
			\ket{\mathrm{i}}\right| ^2\right)\\ 
			&-\frac{\alpha_\mathrm{rot}^2}{4}\mathrm{Re}\left\{\braket{\mathrm{i|f_0}}\bra{\mathrm{f_0}}\hat{\Pi}_\mathrm{j}	\ket{\mathrm{i}}\right\}+\mathcal{O}(\alpha_\mathrm{rot}^3)\\
			=&\left| \braket{\mathrm{f|i}}\right| ^2+\alpha_\mathrm{rot}\mathrm{Im}\left\{ \left|\braket{\mathrm{f_0|i}}\right|^2\frac{\bra{\mathrm{f_0}}\hat{\sigma}^\mathrm{DC}_\mathrm{x}\hat{\Pi}_\mathrm{j}\ket{\mathrm{i}}}{\braket{\mathrm{f_0|i}}}\right\}+\frac{\alpha_\mathrm{rot}^2}{4}\left| \braket{\mathrm{f|i}}\right| ^2\left(\frac{\left| \bra{\mathrm{f_0}}\hat{\sigma}^\mathrm{DC}_\mathrm{x}\hat{\Pi}_j
				\ket{\mathrm{i}}\right| ^2}{\left| \braket{\mathrm{f|i}}\right| ^2}+\frac{\left| \bra{\mathrm{f'}}\hat{\sigma}^\mathrm{DC}_\mathrm{x}\hat{\Pi}_j
				\ket{\mathrm{i}}\right| ^2}{\left| \braket{\mathrm{f|i}}\right| ^2}\right)\\
			&-\frac{\alpha_\mathrm{rot}^2}{4}\mathrm{Re}\left\{ \left|\braket{\mathrm{f_0|i}}\right|^2\frac{\bra{\mathrm{f_0}}\hat{\Pi}_\mathrm{j}\ket{\mathrm{i}}}{\braket{\mathrm{f_0|i}}}\right\}+\mathcal{O}(\alpha_\mathrm{rot}^3)\\
			=&\left| \braket{\mathrm{f|i}}\right| ^2\bigg[1+\alpha_\mathrm{rot}\mathrm{Im}\left\{ \braket{\hat{\sigma}^\mathrm{DC}_\mathrm{x}\hat{\Pi}_j}^\mathrm{E_0}_\mathrm{w}\right\}
			+\frac{\alpha_\mathrm{rot}^2}{4}\left(\frac{\left| \bra{\mathrm{f_0}}\hat{\sigma}^\mathrm{DC}_\mathrm{x}\hat{\Pi}_j
				\ket{\mathrm{i}}\right| ^2}{\left| \braket{\mathrm{f|i}}\right| ^2}+\frac{\left| \bra{\mathrm{f'}}\hat{\sigma}^\mathrm{DC}_\mathrm{x}\hat{\Pi}_j
				\ket{\mathrm{i}}\right| ^2}{\left| \braket{\mathrm{f|i}}\right| ^2}\right)\\
			&-\frac{\alpha_\mathrm{rot}^2}{4}\mathrm{Re}\left\{\braket{\hat{\Pi}_j}^\mathrm{E_0}_\mathrm{w} \right\} +\mathcal{O}(\alpha_\mathrm{rot}^3)\bigg]\\
			=&\left| \braket{\mathrm{f|i}} \right| ^2 \bigg[1+\alpha_\mathrm{rot}\delta_{j,\mathrm{I}}\sin(2\chi_1)+\frac{\alpha_\mathrm{rot}^2}{4} \left(\delta_{j,\mathrm{I}}-\delta_{j,\mathrm{II}}+\delta_{j,\mathrm{III}}\right)
			\bigg]
			+\mathcal{O}(\alpha_\mathrm{rot}^3),
		\end{split}
	\end{align} 
	with the Kronecker delta $\delta_{i,j}$. Similar steps lead to Eq.\,(\ref{CalcRF}). 
	\\
	The step-by-step calculation of the intensity in the absorber case of Eq.\,(\ref{CalcAbs}) is written as 
	\begin{align}
		\begin{split}
			\label{calcAbsDetailed}
			I_{j}^{\mathrm{Abs}}
			=&\left| \braket{\mathrm{f}|\hat{A}^\mathrm{Abs}_j|\mathrm{i}}\right|^2\\	
			=&\left|\bra{\mathrm{f}}\left[\mathds{1}-\hat{\Pi}_j(1-\sqrt{1-\mathcal{A}})\right]\ket{\mathrm{i}}\right|^2\\
			=&\left|\braket{\mathrm{f}|\mathrm{i}}-(1-\sqrt{1-\mathcal{A}})\braket{\mathrm{f}|\hat{\Pi}_j|\mathrm{i}}\right|^2\\
			=&\left|\braket{\mathrm{f}|\mathrm{i}}\right|^2-(1-\sqrt{1-\mathcal{A}})\left(\braket{\mathrm{f}|\mathrm{i}}\braket{\mathrm{i}|\hat{\Pi}_j|\mathrm{f}}+\braket{\mathrm{i}|\mathrm{f}}\braket{\mathrm{f}|\hat{\Pi}_j|\mathrm{i}}\right)+(1-\sqrt{1-\mathcal{A}})^2\braket{\mathrm{i}|\hat{\Pi}_j|\mathrm{f}}\braket{\mathrm{f}|\hat{\Pi}_j|\mathrm{i}}\\
			=&\left|\braket{\mathrm{f}|\mathrm{i}}\right|^2-2(1-\sqrt{1-\mathcal{A}})\mathrm{Re}\left\{\braket{\mathrm{i}|\mathrm{f}}\braket{\mathrm{f}|\hat{\Pi}_j|\mathrm{i}}\right\}+(1-\sqrt{1-\mathcal{A}})^2\braket{\mathrm{i}|\hat{\Pi}_j|\mathrm{f}}\braket{\mathrm{f}|\hat{\Pi}_j|\mathrm{i}}\\
			=&\left|\braket{\mathrm{f}|\mathrm{i}}\right|^2-2(1-\sqrt{1-\mathcal{A}})\mathrm{Re}\left\{\braket{\mathrm{i}|\mathrm{f}}\left(\ket{\mathrm{E_0}}\bra{\mathrm{E_0}}+\ket{\mathrm{E'}}\bra{\mathrm{E'}}\right)\braket{\mathrm{f}|\hat{\Pi}_j|\mathrm{i}}\right\}\\
			&+(1-\sqrt{1-\mathcal{A}})^2\braket{\mathrm{i}|\hat{\Pi}_j|\mathrm{f}}\left(\ket{\mathrm{E_0}}\bra{\mathrm{E_0}}+\ket{\mathrm{E'}}\bra{\mathrm{E'}}\right)\braket{\mathrm{f}|\hat{\Pi}_j|\mathrm{i}}\\
			=&\left|\braket{\mathrm{f}|\mathrm{i}}\right|^2-2(1-\sqrt{1-\mathcal{A}})\mathrm{Re}\left\{\braket{\mathrm{i}|\mathrm{f_0}}\braket{\mathrm{f_0}|\hat{\Pi}_j|\mathrm{i}}\right\}+(1-\sqrt{1-\mathcal{A}})^2\braket{\mathrm{i}|\hat{\Pi}_j|\mathrm{f_0}}\braket{\mathrm{f_0}|\hat{\Pi}_j|\mathrm{i}}\\
			=&\left|\braket{\mathrm{f}|\mathrm{i}}\right|^2-2(1-\sqrt{1-\mathcal{A}})\mathrm{Re}\left\{\left|\braket{\mathrm{f_0|i}}\right|^2\frac{\bra{\mathrm{f_0}}\hat{\Pi}_\mathrm{j}\ket{\mathrm{i}}}{\braket{\mathrm{f_0|i}}}\right\}+(1-\sqrt{1-\mathcal{A}})^2\left|\braket{\mathrm{f_0|i}}\right|^2\left|\frac{ \bra{\mathrm{f_0}}\hat{\Pi}_j
				\ket{\mathrm{i}}}{\braket{\mathrm{f_0|i}}}\right| ^2\\
			=&\left| \braket{\mathrm{f|i}} \right| ^2 \left[1-2(1-\sqrt{1-\mathcal{A}})\mathrm{Re}\left\{ \braket{\hat{\Pi}_j}^\mathrm{E_0}_\mathrm{w}\right\}+(1-\sqrt{1-\mathcal{A}})^2\left|\braket{\hat{\Pi}_j}^\mathrm{E_0}_\mathrm{w}\right| ^2\right]\\
			=&\left| \braket{\mathrm{f|i}} \right| ^2 \left[1-2(1-\sqrt{1-\mathcal{A}})\braket{\hat{\Pi}_j}^\mathrm{E_0}_\mathrm{w}
			+(1-\sqrt{1-\mathcal{A}})^2\braket{\hat{\Pi}_j}^\mathrm{E_0}_\mathrm{w}\right]\\
			=&\left| \braket{\mathrm{f|i}} \right| ^2 \left[1-\bigg(2(1-\sqrt{1-\mathcal{A}})
			-(1-\sqrt{1-\mathcal{A}})^2\bigg)\braket{\hat{\Pi}_j}^\mathrm{E_0}_\mathrm{w}\right]\\
			=&\left| \braket{\mathrm{f|i}} \right| ^2 \left[1-\mathcal{A}\braket{\hat{\Pi}_j}^\mathrm{E_0}_\mathrm{w}
			\right]\\
			=&\left| \braket{\mathrm{f|i}} \right| ^2 \left[1-\mathcal{A}\,\delta_{j,\mathrm{II}} \right], 
		\end{split}
	\end{align} 
	where it is used that, according to Eq.\,(\ref{weakValueCalc}),
	\begin{equation}
		\label{RelationPathWeakValue}
		\braket{\hat{\Pi}_j}^\mathrm{E_0}_\mathrm{w}=\delta_{j,\mathrm{II}}=\mathrm{Re}\left\{ \braket{\hat{\Pi}_j}^\mathrm{E_0}_\mathrm{w}\right\}=\left|\braket{\hat{\Pi}_j}^\mathrm{E_0}_\mathrm{w}\right| ^2.
	\end{equation}
	
	The detailed calculation of the exact result with N paths and N-1 properties of Eq.\,(\ref{CalcAlok}) is given here: 
	\begin{align}
		\begin{split}
			\label{calcAlokDetailed}
			I_j^{p}(\alpha)=&\left| \bra{\mathrm{f_N}}\hat{O}^p_j(\alpha)\mathrm{\ket{i_N}}\right|^2\\	
			=&\left|\bra{\mathrm{f_N}}\left[\mathds{1}-\left(1-\cos\frac{\alpha}{2}\right)\hat{\Pi}_j-\mathrm{i}\sin\left(\frac{\alpha}{2}\right)\hat{\sigma}^p_\mathrm{x}\hat{\Pi}_j\right]\ket{\mathrm{i_N}}\right|^2\\
			=&\left| \braket{\mathrm{f_N|i_N}}-\left(1-\cos\frac{\alpha}{2}\right)\bra{\mathrm{f_N}}\hat{\Pi}_j\ket{\mathrm{i_N}}
			-\mathrm{i}\sin\left(\frac{\alpha}{2}\right)\bra{\mathrm{f_N}}\hat{\sigma}^p_\mathrm{x}\hat{\Pi}_j
			\ket{\mathrm{i_N}}\right| ^2\\
			=&\left| \braket{\mathrm{f_N|i_N}}\right|^2+\left(1-\cos\frac{\alpha}{2}\right)^2\left|\bra{\mathrm{f_N}}\hat{\Pi}_j\ket{\mathrm{i_N}}\right|^2+\sin^2\left(\frac{\alpha}{2}\right)\left|\bra{\mathrm{f_N}}\hat{\sigma}^p_\mathrm{x}\hat{\Pi}_j
			\ket{\mathrm{i_N}}\right| ^2\\
			&-\braket{\mathrm{i_N|f_N}}\left(1-\cos\frac{\alpha}{2}\right)\bra{\mathrm{f_N}}\hat{\Pi}_j\ket{\mathrm{i_N}}-\braket{\mathrm{f_N|i_N}}\left(1-\cos\frac{\alpha}{2}\right)\bra{\mathrm{i_N}}\hat{\Pi}_j\ket{\mathrm{f_N}}\\
			&-\braket{\mathrm{i_N|f_N}}\mathrm{i}\sin\left(\frac{\alpha}{2}\right)\bra{\mathrm{f_N}}\hat{\sigma}^p_\mathrm{x}\hat{\Pi}_j\ket{\mathrm{i_N}}+\braket{\mathrm{f_N|i_N}}\mathrm{i}\sin\left(\frac{\alpha}{2}\right)\bra{\mathrm{i_N}}\hat{\sigma}^p_\mathrm{x}\hat{\Pi}_j\ket{\mathrm{f_N}}\\
			&+\left(1-\cos\frac{\alpha}{2}\right)\braket{\mathrm{i_N|\hat{\Pi}_j|f_N}}\mathrm{i}\sin\left(\frac{\alpha}{2}\right)\bra{\mathrm{f_N}}\hat{\sigma}^p_\mathrm{x}\hat{\Pi}_j\ket{\mathrm{i_N}}
			-\left(1-\cos\frac{\alpha}{2}\right)\braket{\mathrm{f_N|\hat{\Pi}_j|i_N}}\mathrm{i}\sin\left(\frac{\alpha}{2}\right)\bra{\mathrm{i_N}}\hat{\sigma}^p_\mathrm{x}\hat{\Pi}_j\ket{\mathrm{f_N}},
		\end{split}
	\end{align} 
	where the last two summands are zero because at least one of the two brakets in each term is zero. This is in close relation to the calculation of the weak values in Eq.\,(\ref{weakValueCalc}) but not exemplified here. Continuing the derivation,  
	\begin{align}
		\begin{split}
			I_j^{p}(\alpha)=&\left| \braket{\mathrm{f_N|i_N}}\right|^2\bigg[ 1+\left(1-\cos\frac{\alpha}{2}\right)^2\left|\frac{\bra{\mathrm{f_N}}\hat{\Pi}_j\ket{\mathrm{i_N}}}{\braket{\mathrm{f_N|i_N}}}\right|^2+\sin^2\left(\frac{\alpha}{2}\right)\left|\frac{\bra{\mathrm{f_N}}\hat{\sigma}^p_\mathrm{x}\hat{\Pi}_j
				\ket{\mathrm{i_N}}}{\braket{\mathrm{f_N|i_N}}}\right| ^2\\
			&-\left(1-\cos\frac{\alpha}{2}\right)\frac{\bra{\mathrm{f_N}}\hat{\Pi}_j\ket{\mathrm{i_N}}}{\braket{\mathrm{f_N|i_N}}}-\left(1-\cos\frac{\alpha}{2}\right)\frac{\bra{\mathrm{i_N}}\hat{\Pi}_j\ket{\mathrm{f_N}}}{\braket{\mathrm{i_N|f_N}}}\\
			&-\mathrm{i}\sin\left(\frac{\alpha}{2}\right)\frac{\bra{\mathrm{f_N}}\hat{\sigma}^p_\mathrm{x}\hat{\Pi}_j
				\ket{\mathrm{i_N}}}{\braket{\mathrm{f_N|i_N}}}+\mathrm{i}\sin\left(\frac{\alpha}{2}\right)\frac{\bra{\mathrm{i_N}}\hat{\sigma}^p_\mathrm{x}\hat{\Pi}_j
				\ket{\mathrm{f_N}}}{\braket{\mathrm{i_N|f_N}}}\bigg]\\
			=&\left| \braket{\mathrm{f_N|i_N}}\right|^2\bigg[ 1+\left(1-\cos\frac{\alpha}{2}\right)^2\left|\braket{\hat{\Pi}_j}_\mathrm{w}\right|^2+\sin^2\left(\frac{\alpha}{2}\right)\left|\braket{\hat{\sigma}^p_\mathrm{x}\hat{\Pi}_j}_\mathrm{w}\right| ^2\\
			&-\left(1-\cos\frac{\alpha}{2}\right)\left(\braket{\hat{\Pi}_j}_\mathrm{w}+\braket{\hat{\Pi}_j}_\mathrm{w}^*\right)
			-\mathrm{i}\sin\left(\frac{\alpha}{2}\right)\left(\braket{\hat{\sigma}^p_\mathrm{x}\hat{\Pi}_j}_\mathrm{w}-\braket{\hat{\sigma}^p_\mathrm{x}\hat{\Pi}_j}_\mathrm{w}^*\right)\bigg]\\
			=&\left| \braket{\mathrm{f_N|i_N}}\right|^2\bigg[ 1+\left(1-\cos\frac{\alpha}{2}\right)^2\left|\braket{\hat{\Pi}_j}_\mathrm{w}\right|^2+\sin^2\left(\frac{\alpha}{2}\right)\left|\braket{\hat{\sigma}^p_\mathrm{x}\hat{\Pi}_j}_\mathrm{w}\right| ^2\\
			&-2\left(1-\cos\frac{\alpha}{2}\right)\mathrm{Re}\left\{\braket{\hat{\Pi}_j}_\mathrm{w}\right\}
			-2\mathrm{i^2}\sin\left(\frac{\alpha}{2}\right)\mathrm{Im}\left\{\braket{\hat{\sigma}^p_\mathrm{x}\hat{\Pi}_j}_\mathrm{w}\right\}\bigg]\\
			=&\left| \braket{\mathrm{f_N|i_N}}\right|^2\bigg[ 1+2\sin\left(\frac{\alpha}{2}\right)\mathrm{Im}\left\{\braket{\hat{\sigma}^p_\mathrm{x}\hat{\Pi}_j}_\mathrm{w}\right\}+\sin^2\left(\frac{\alpha}{2}\right)\left|\braket{\hat{\sigma}^p_\mathrm{x}\hat{\Pi}_j}_\mathrm{w}\right| ^2\\
			&+\left(\left(1-\cos\frac{\alpha}{2}\right)^2-2\left(1-\cos\frac{\alpha}{2}\right)\right)\braket{\hat{\Pi}_j}_\mathrm{w}
			\bigg]\\
			=&\left| \braket{\mathrm{f_N|i_N}}\right|^2\bigg[ 1+2\sin\left(\frac{\alpha}{2}\right)\mathrm{Im}\left\{\braket{\hat{\sigma}^p_\mathrm{x}\hat{\Pi}_j}_\mathrm{w}\right\}+\sin^2\left(\frac{\alpha}{2}\right)\left|\braket{\hat{\sigma}^p_\mathrm{x}\hat{\Pi}_j}_\mathrm{w}\right| ^2-\sin^2\left(\frac{\alpha}{2}\right)\braket{\hat{\Pi}_j}_\mathrm{w}\bigg],
		\end{split}
	\end{align} 
	where we also use Eq.\,(\ref{RelationPathWeakValue}). This concludes the derivation of the second line in Eq.\,(\ref{CalcAlok}). The derivation of lines three and four of Eq.\,(\ref{CalcAlok}) takes a different way. By explicitly writing the brakets, given by Eqs.\,(\ref{PreselectionAlok},\ref{PostselectionAlok}), in the third line of Eq.\,(\ref{calcAlokDetailed}), we obtain
	
	\begin{align}
		\begin{split}
			I_j^{p}(\alpha)=&\bigg| \braket{\mathrm{f_N|i_N}}-\left(1-\cos\frac{\alpha}{2}\right)\bigg[\frac{1}{\sqrt{\mathrm{N}}}\left(\mathrm{e}^{-\mathrm{i}\chi_1}\bra{\mathrm{I}}+\mathrm{e}^{-\mathrm{i}\chi_2}\bra{\mathrm{II}}+...+\mathrm{e}^{-\mathrm{i}\chi_N}\bra{\mathrm{N}}\right) \bra{1,1,...1,1}\bigg]\hat{\Pi}_j\times\\
			&\times\bigg[\frac{1}{\sqrt{\mathrm{N}}}\big( \ket{\mathrm{I}}\ket{1,1,...1,1}+\ket{\mathrm{II}}\ket{0,1,...1,1}+...+\ket{\mathrm{N}}\ket{1,1,...1,0}\big)\bigg]\\
			&-\mathrm{i}\sin\frac{\alpha}{2}\bigg[\frac{1}{\sqrt{\mathrm{N}}}\left(\mathrm{e}^{-\mathrm{i}\chi_1}\bra{\mathrm{I}}+\mathrm{e}^{-\mathrm{i}\chi_2}\bra{\mathrm{II}}+...+\mathrm{e}^{-\mathrm{i}\chi_N}\bra{\mathrm{N}}\right) \bra{1,1,...1,1}\bigg]\hat{\sigma}^p_\mathrm{x}\hat{\Pi}_j\times\\
			&\times\bigg[\frac{1}{\sqrt{\mathrm{N}}}\big( \ket{\mathrm{I}}\ket{1,1,...1,1}+\ket{\mathrm{II}}\ket{0,1,...1,1}+...+\ket{\mathrm{N}}\ket{1,1,...1,0}\big)\bigg]\bigg| ^2\\
			=&\left| \frac{1}{\mathrm{N}}\mathrm{e}^{-\mathrm{i}\chi_1}-\left(1-\cos\frac{\alpha}{2}\right)\frac{1}{\mathrm{N}}\mathrm{e}^{-\mathrm{i}\chi_1}\delta_{j,1}
			-\mathrm{i}\sin\left(\frac{\alpha}{2}\right)\frac{1}{\mathrm{N}}\mathrm{e}^{-\mathrm{i}\chi_p}\delta_{j,p+1}\right| ^2\\
			=& \frac{1}{\mathrm{N^2}}\bigg[ 1+\left(1-\cos\frac{\alpha}{2}\right)^2 \delta_{j,1}
			+\sin^2\left(\frac{\alpha}{2}\right)\delta_{j,p+1}\\
			&-\left(1-\cos\frac{\alpha}{2}\right)\mathrm{e}^{-\mathrm{i}\chi_1}\delta_{j,1}\mathrm{e}^{\mathrm{i}\chi_1}-\left(1-\cos\frac{\alpha}{2}\right)\mathrm{e}^{\mathrm{i}\chi_1}\delta_{j,1}\mathrm{e}^{-\mathrm{i}\chi_1}\\
			&-\mathrm{i}\sin\left(\frac{\alpha}{2}\right)\mathrm{e}^{\mathrm{i}(\chi_1-\chi_p)}\delta_{j,p+1}+\mathrm{i}\sin\left(\frac{\alpha}{2}\right)\mathrm{e}^{-\mathrm{i}(\chi_1-\chi_p)}\delta_{j,p+1}\\
			&-\left(1-\cos\frac{\alpha}{2}\right)\mathrm{e}^{-\mathrm{i}\chi_1}\delta_{j,1}\mathrm{i}\sin\left(\frac{\alpha}{2}\right)\mathrm{e}^{\mathrm{i}\chi_p}\delta_{j,p+1}-\left(1-\cos\frac{\alpha}{2}\right)\delta_{j,1}(-\mathrm{i})\sin\left(\frac{\alpha}{2}\right)\mathrm{e}^{-\mathrm{i}(\chi_p-\chi_\mathrm{1})}\delta_{j,p+1}\bigg]\\	
			=&\frac{1}{\mathrm{N^2}}\bigg[ 1-\delta_{j,p+1}\mathrm{i}\sin\left(\frac{\alpha}{2}\right)\left(\mathrm{e}^{\mathrm{i}(\chi_1-\chi_p)}-\mathrm{e}^{-\mathrm{i}(\chi_1-\chi_p)}\right)+\delta_{j,p+1}\sin^2\left(\frac{\alpha}{2}\right)\\
			&+\delta_{j,1}\left(\left(1-\cos\frac{\alpha}{2}\right)^2-2\left(1-\cos\frac{\alpha}{2}\right)\right)\bigg]\\		
			=&\frac{1}{\mathrm{N^2}}\bigg[ 1-\delta_{j,p+1}\mathrm{i}\sin\left(\frac{\alpha}{2}\right)\frac{2\mathrm{i}}{2\mathrm{i}}\left(\mathrm{e}^{\mathrm{i}(\chi_1-\chi_p)}-\mathrm{e}^{-\mathrm{i}(\chi_1-\chi_p)}\right)+\delta_{j,p+1}\sin^2\left(\frac{\alpha}{2}\right)       +\delta_{j,1}\left[\cos^2\left(\frac{\alpha}{2}\right)-1\right]\bigg]	\\
			=&\frac{1}{\mathrm{N^2}}\bigg[ 1+2\delta_{j,p+1}\sin\left(\frac{\alpha}{2}\right)\sin(\chi_1-\chi_p)+\delta_{j,p+1}\sin^2\left(\frac{\alpha}{2}\right)-\delta_{j,1}\sin^2\left(\frac{\alpha}{2}\right)
			\bigg].	
		\end{split}
	\end{align} 
	
	\subsection{Extraction of Weak Values}
	\label{DetailedDataAnalysis}
	
	Parameters in this section with indices “empty”, “prep” and “weak” correspond to the three types of interferograms recorded, respectively, with either no elements in the interferometer, with the preparational DC and RF flip applied, and an additional weak interaction applied. To read out the signal generated by the weak interaction, we assume that, in the case of the weak interaction IFGs, the intensity oscillation of Eq.\,(\ref{fitfunction}) is the sum of two independent oscillations:
	\begin{align}
		\begin{split}
			\label{AnsatzWV}
			I_{\mathrm{weak}}(\chi)=&I_{\mathrm{0,weak}}+B_{\mathrm{weak}}\cos(\omega_{\mathrm{empty}}\chi+\varphi_{\mathrm{weak}})\\ =&I_{\mathrm{0,weak}}+B_{\mathrm{prep}}\cos(\omega_{\mathrm{empty}}\chi+\varphi_{\mathrm{prep}})+B_{\mathrm{signal}}\cos(\omega_{\mathrm{empty}}\chi+\varphi_{\mathrm{signal}}).
		\end{split}
	\end{align}
	\\
	Therein, the “signal” refers to the changes in the IFGs from the preparational case to the weak interaction case. The amplitude and phase of the signal can be retrieved by comparing interferograms of the sets consisting of an interferogram with only the preparation applied and an interferogram with an additional weak interaction applied.
	The signal amplitude $B_{\mathrm{signal}}$ and its statistical error $\Delta B_{\mathrm{signal}}$ follow from Eq.\,(\ref{AnsatzWV}) as 
	\begin{align}
		B_{\mathrm{signal}}=\sqrt{B_{\mathrm{weak}}^2+B_{\mathrm{prep}}^2-2B_{\mathrm{weak}}B_{\mathrm{prep}}\cos(\varphi_{\mathrm{weak}}-\varphi_{\mathrm{prep}})}
	\end{align}
	and
	\begin{align}
		\begin{split}
			\label{errorPropagation}
			\Delta B_{\mathrm{signal}}=\frac{1}{B_{\mathrm{signal}}}\bigg[&\left[\left(B_{\mathrm{weak}}-B_{\mathrm{prep}}\cos(\varphi_{\mathrm{weak}}-\varphi_{\mathrm{prep}})\right)\Delta B_{\mathrm{weak}}\right]^2\\
			+&\left[\left(B_{\mathrm{prep}}-B_{\mathrm{weak}}\cos(\varphi_{\mathrm{weak}}-\varphi_{\mathrm{prep}})\right)\Delta B_{\mathrm{prep}}\right]^2\\
			+&\left(B_{\mathrm{weak}} B_{\mathrm{prep}}\sin(\varphi_{\mathrm{weak}}-\varphi_{\mathrm{prep}})\right)^2(\Delta \varphi_{\mathrm{weak}}^2+\Delta \varphi_{\mathrm{prep}}^2)
			\bigg]^{1/2}.
		\end{split}
	\end{align}
	\\	
	The weak values are extracted by comparing experimental data with the theoretical prediction. By substituting the second last equality in Eq.\,(\ref{calcDCdetailed}) into Eq.\,(\ref{AnsatzWV}) and neglecting terms of order higher than $\alpha_\mathrm{rot}$ we obtain 
	\begin{equation}
		\left| \braket{\mathrm{f|i}}\right| ^2\bigg[1+C_{\mathrm{empty}}\alpha_\mathrm{rot}\mathrm{Im}\left\{ \braket{\hat{\sigma}^\mathrm{DC}_\mathrm{x}\hat{\Pi}}^\mathrm{E_0}_\mathrm{w}\right\}\bigg]=I_{\mathrm{0,weak}}+B_{\mathrm{prep}}\cos(\omega_{\mathrm{empty}}\chi+\varphi_{\mathrm{prep}})+B_{\mathrm{signal}}\cos(\omega_{\mathrm{empty}}\chi+\varphi_{\mathrm{signal}}),
	\end{equation}
	with the correction considering the maximum experimental contrast $C_{\mathrm{empty}}$ of the empty interferometer given through the fits in Fig.\,\ref{IFGsEmpty}. The index $j$ of the path where the rotation is implemented is omitted here. We can drop the oscillation proportional to $B_{\mathrm{prep}}$ already present in the preparational IFGs as it is an experimental imperfection and does not represent the behaviour described by weak values. It is however included through the error propagation of Eq.\,(\ref{errorPropagation}). Furthermore, we can insert $\left| \braket{\mathrm{f|i}}\right| ^2=I_{0,\mathrm{prep}}\approx I_{0,\mathrm{weak}}$. In this context, it is important to discern between the phase $\chi'$ of the wave function and the phase shifter position $\chi$ that are related via $\chi'=\omega_{\mathrm{empty}}\chi+const.$ such that $\omega_{\mathrm{empty}}\chi+\varphi_{\mathrm{signal}}=\chi'+\varphi'_{\mathrm{signal}}$. It follows that
	\begin{align}
		\begin{split}
			I_{0,\mathrm{prep}} \bigg[1+C_{\mathrm{empty}}\alpha_\mathrm{rot}\mathrm{Im}\left\{ \braket{\hat{\sigma}^\mathrm{DC}_\mathrm{x}\hat{\Pi}}^\mathrm{E_0}_\mathrm{w}\right\}\bigg]&=I_{\mathrm{0,weak}}+B_{\mathrm{signal}}\cos(\chi'+\varphi'_{\mathrm{signal}}),\\
			1+C_{\mathrm{empty}}\alpha_\mathrm{rot}\mathrm{Im}\left\{ \braket{\hat{\sigma}^\mathrm{DC}_\mathrm{x}\hat{\Pi}}^\mathrm{E_0}_\mathrm{w}\right\}&=\frac{I_{\mathrm{0,weak}}}{I_{0,\mathrm{prep}}}+\frac{B_{\mathrm{signal}}}{I_{0,\mathrm{prep}}}\cos(\chi'+\varphi'_{\mathrm{signal}}),\\
			C_{\mathrm{empty}}\alpha_\mathrm{rot}\mathrm{Im}\left\{ \braket{\hat{\sigma}^\mathrm{DC}_\mathrm{x}\hat{\Pi}}^\mathrm{E_0}_\mathrm{w}\right\}&\approx\frac{B_{\mathrm{signal}}}{I_{0,\mathrm{prep}}}\cos(\chi'+\varphi'_{\mathrm{signal}}),\\
			\frac{\mathrm{Im}\left\{ \braket{\hat{\sigma}^\mathrm{DC}_\mathrm{x}\hat{\Pi}}^\mathrm{E_0}_\mathrm{w}\right\}}{\cos(\chi'+\varphi'_{\mathrm{signal}})}&\approx\frac{\frac{B_{\mathrm{signal}}}{I_{0,\mathrm{prep}}}}{C_{\mathrm{empty}}\alpha_\mathrm{rot}},\ \ \ \forall\ \chi'\in \mathbb{R}.
		\end{split}
	\end{align}
	Since this relation must hold for all $\chi'$, the imaginary part of the weak value must be sinusoidal as obtained in Eq.\,(\ref{weakValueCalc}). Furthermore, the cosine function and the imaginary part of the weak value must have the same frequency and be in phase. The weak values of Eq.\,(\ref{weakValueCalc}) all have constant absolute values and we also assume this to hold for all extracted weak values. Thus we finally obtain in first order of $\alpha_\mathrm{rot}$ the measured absolute value of the weak value 
	\begin{align}
		\left|\braket{\hat{\sigma}^\mathrm{DC}_\mathrm{x}\hat{\Pi}}^\mathrm{E_0}_{\mathrm{w}}\right| = \frac{\frac{B_{\mathrm{signal}}}{I_{0,\mathrm{prep}}}}{C_{\mathrm{empty}}\alpha_\mathrm{rot}}, 
	\end{align}
	and its statistical error
	\begin{align}
		\Delta\left|\braket{\hat{\sigma}^\mathrm{DC}_\mathrm{x}\hat{\Pi}}^\mathrm{E_0}_{\mathrm{w}}\right|=\frac{B_{\mathrm{signal}}}{I_{0,\mathrm{prep}}C_{\mathrm{empty}}\alpha_\mathrm{rot}} \sqrt{ \left(\frac{\Delta B_{\mathrm{signal}}}{B_{\mathrm{signal}}}\right)^2+\left(\frac{\Delta I_{0,\mathrm{prep}}}{I_{0,\mathrm{prep}}}\right)^2+\left(\frac{\Delta C_{\mathrm{empty}}}{C_{\mathrm{empty}}}\right)^2+\left(\frac{\Delta\alpha_\mathrm{rot}}{\alpha_\mathrm{rot}}\right)^2}. 
	\end{align}
	The same steps lead to a similar result for the RF case. For the case of weak absorption, we measured the absorption coefficient of the Indium foil with a single interferometer path as
	\begin{equation}
		\mathcal{A}=1-0.90(1)=0.10(1).\\
	\end{equation}
	
	We substitute the second last line in Eq.\,(\ref{calcAbsDetailed}) into Eq.\,(\ref{AnsatzWV}) such that 
	\begin{equation}
		\left| \braket{\mathrm{f|i}} \right| ^2 \left[1-\braket{\hat{\Pi}}^\mathrm{E_0}_\mathrm{w}\mathcal{A}
		\right]=I_{\mathrm{0,weak}}+A_{\mathrm{prep}}\cos(\omega_{\mathrm{empty}}\chi+\varphi_{\mathrm{prep}})+A_{\mathrm{signal}}\cos(\omega_{\mathrm{empty}}\chi+\varphi_{\mathrm{signal}}).
	\end{equation}
	The index $j$ of the path where the absorption is implemented is omitted again. Both oscillations can be neglected as they neither describe a consequence to the weak absorption nor change the mean intensity. With similar steps as for the DC case we calculate 
	\begin{align}
		\begin{split}
			\left| \braket{\mathrm{f|i}} \right| ^2 \left[1-\braket{\hat{\Pi}}^\mathrm{E_0}_\mathrm{w}\mathcal{A}\right]&=I_{\mathrm{0,weak}}\\
			I_{\mathrm{0,prep}}\left[1-\braket{\hat{\Pi}}^\mathrm{E_0}_\mathrm{w}\mathcal{A}\right]&=I_{\mathrm{0,weak}}\\
			1-\braket{\hat{\Pi}}^\mathrm{E_0}_\mathrm{w}\mathcal{A}&=\frac{I_{\mathrm{0,weak}}}{I_{\mathrm{0,prep}}}\\
			\braket{\hat{\Pi}}^\mathrm{E_0}_\mathrm{w}\mathcal{A}&=1-\frac{I_{\mathrm{0,weak}}}{I_{\mathrm{0,prep}}}\\
			\braket{\hat{\Pi}}^\mathrm{E_0}_\mathrm{w}&=\frac{\mathcal{A}_\mathrm{w}}{\mathcal{A}},\\
		\end{split}
	\end{align}
	with the effective absorption coefficient $\mathcal{A}_\mathrm{w}$ in the path of the interferometer where the absorber is inserted written as
	\begin{equation}
		\mathcal{A}_\mathrm{w}=1-\frac{I_{\mathrm{0,weak}}}{I_{\mathrm{0,prep}}}.
	\end{equation}
	The propagated statistical error of the path weak value is given by
	\begin{equation}
		\Delta\braket{\hat{\Pi}}^\mathrm{E_0}_\mathrm{w} = \frac{1}{\mathcal{A}} \sqrt{\left[\left(1-\frac{I_{\mathrm{0,weak}}}{I_{\mathrm{0,prep}}}\right) \frac{\Delta\mathcal{A}}{\mathcal{A}}\right]^2 
			+ \left(I_{\mathrm{0,weak}} \frac{\Delta I_{\mathrm{0,prep}}}{
				I_{\mathrm{0,prep}}^2}\right)^2 + \left(\frac{\Delta I_{\mathrm{0,weak}}}{I_{\mathrm{0,prep}}}\right)^2}.
	\end{equation}

	\emph{Data Availability.---}%
	The data that support the findings of this study are available under http://doi.ill.fr/10.5291/ILL-DATA.3-16-8 . 
	
	\emph{Acknowledgements.---}%
	We are grateful to Michael Jentschel for his courteous aid in the preparation of the water cooling system and the ILL for its continued hospitality. We thank Alok Kumar Pan and Andreas Dvorak for fruitful discussion, and Erwin Jericha for his critical response concerning the terminology of properties. This work was financed by
	Austrian science fund (FWF) Projects No. P 30677-N36 and P 27666-N20. Y.H. is partly supported by KAKENHI Project No. 18H03466.

	\emph{Author contributions.---}%
	A.D., K.O., R.W., and Y.H. conceived the experiment, A.D., K.O., and H.L. prepared it, A.D., N.G., H.L., and R.W. conducted it, A.D. and Y.H. analysed the data, all authors co-wrote the paper.

	\emph{Additional Information.---}%
	Competing financial interests: The authors declare no competing financial interests.

	\bibliographystyle{unsrt}
	\typeout{}
	\bibliography{bibliography}
	
\end{document}